\documentclass{article}

 \usepackage[accepted]{icml2025}

\usepackage{amsmath,amsfonts,bm}

\def\figref#1{figure~\ref{#1}}
\def\Figref#1{Figure~\ref{#1}}

\def\secref#1{section~\ref{#1}}

\def\eqref#1{equation~\ref{#1}}

\def\1{\bm{1}}

\DeclareMathAlphabet{\mathsfit}{\encodingdefault}{\sfdefault}{m}{sl}
\SetMathAlphabet{\mathsfit}{bold}{\encodingdefault}{\sfdefault}{bx}{n}

\DeclareMathOperator*{\argmin}{arg\,min}

\usepackage{hyperref}
\usepackage{url}

\usepackage[utf8]{inputenc} 
\usepackage[T1]{fontenc}    
\usepackage{booktabs}       
\usepackage{amsfonts}       
\usepackage{nicefrac}       
\usepackage{microtype}      
\usepackage{microtype}
\usepackage{graphicx}
\usepackage{booktabs} 
\usepackage{amsmath}
\usepackage{amssymb}
\usepackage{mathtools}
\usepackage{amsthm}

\usepackage[capitalize,noabbrev]{cleveref}

\usepackage{comment}
\usepackage{multirow}
\usepackage{xcolor}
\usepackage{algorithm}
\usepackage{algorithmic}
\usepackage{soul}
\usepackage{xspace}
\usepackage{wrapfig}
\usepackage{nicefrac}
\usepackage{tabularx}
\usepackage{bm}
\usepackage{upgreek}
\usepackage{subcaption}

\usepackage{pifont}
\newcommand{\cmark}{\ding{51}}%
\newcommand{\xmark}{\ding{55}}%

\newcommand{\dr}{\delta\mathbf{r}}%

\let\oldAA\AA
\renewcommand{\AA}{\text{\normalfont\oldAA}}

\newcommand{\tabref}[1]{Table~\ref{#1}}
\newcommand{\Tabref}[1]{Table~\ref{#1}}

\renewcommand{\eqref}[1]{Eq.~(\ref{#1})}

\icmltitlerunning{Physics-Informed Weakly Supervised Learning for Interatomic Potentials}

\begin{document}

\twocolumn[
\icmltitle{Physics-Informed Weakly Supervised Learning for Interatomic Potentials}



\icmlsetsymbol{equal}{*}

\begin{icmlauthorlist}
\icmlauthor{Makoto Takamoto}{nle}
\icmlauthor{Viktor Zaverkin}{nle}
\icmlauthor{Mathias Niepert}{nle,us}

\end{icmlauthorlist}

\icmlaffiliation{nle}{NEC Laboratories Europe, Heidelberg, Germany}
\icmlaffiliation{us}{University of Stuttgart, Stuttgart, Germany}

\icmlcorrespondingauthor{Makoto Takamoto}{makoto.takamoto@neclab.eu}

\icmlkeywords{Interatomic Potential Model, Physics Informed method, Weakly Supervised Approach, Machine Learning}

\vskip 0.3in
]



\printAffiliationsAndNotice{}  

\begin{abstract}
    Machine learning plays an increasingly important role in computational chemistry and materials science, complementing computationally intensive ab initio and first-principles methods. Despite their utility, machine-learning models often lack generalization capability and robustness during atomistic simulations, yielding unphysical energy and force predictions that hinder their real-world applications. We address this challenge by introducing a physics-informed, weakly supervised approach for training machine-learned interatomic potentials (MLIPs). We introduce two novel loss functions, extrapolating the potential energy via a Taylor expansion and using the concept of conservative forces. Our approach improves the accuracy of MLIPs applied to training tasks with sparse training data sets and reduces the need for pre-training computationally demanding models with large data sets. Particularly, we perform extensive experiments demonstrating reduced energy and force errors---often lower by a factor of two---for various baseline models and benchmark data sets. Moreover, we demonstrate improved robustness during MD simulations of the MLIP models trained with the proposed weakly supervised loss. Finally, our approach improves the fine-tuning of foundation models on sparse, highly accurate ab initio data. An implementation of our method and scripts for executing experiments are available at \url{https://github.com/nec-research/PICPS-ML4Sci}.
\end{abstract}

\section{Introduction \label{sec:introduction}}

Ab initio and first-principles methods are inevitable for the computer-aided exploration of molecular and material properties used in the chemical sciences and engineering~\citep{Parrinello1997, Carloni2002, Iftimie2005}. However, commonly employed ab initio and first-principles approaches---such as coupled cluster (CC) \citep{Purvis1982, Bartlett2007} and density functional theory (DFT) \citep{Hohenberg1964, Kohn1965}, respectively---require substantial compute resources. Thus, they typically allow only for atomistic simulations of small- to medium-sized atomic systems and restrict the accessible simulation times, which affects the accuracy of estimated molecular and material properties. Classical force fields can extend these length and time scales, providing a computationally efficient alternative to first-principles approaches, but often lack accuracy. Machine-learning-based models hold promise to bridge the gap between first-principles and classical approaches, yielding computationally efficient and accurate machine-learned interatomic potentials (MLIPs)~\citep{Smith2017, ocp_dataset, Unke2021, Merchant2023, Kovacs2023, Batatia2023}. 

These MLIPs, however, face several challenges. They require the generation of training data sets that sufficiently cover configurational (atom positions) and compositional (atom types) spaces using, e.g., molecular dynamics (MD) simulations based on ab initio or first-principles approaches. Given the high computational cost of the commonly used data generation approaches, the resulting training data sets are often sparse and prevent the application of MLIPs to new molecular and material systems. Active learning can be used to address this challenge~\citep{Li2015, Vandermause2020, Zaverkin2021a, Zaverkin2022d, vanerOord2022, zaverkin2024uncertainty}, but still requires generating a non-negligible number of first-principles (e.g., DFT) data to train the initial model, which is then used to explore the phase space with sufficiently long MD simulations. Hence, a strong motivation for the proposed method is its use in combination with active learning to acquire additional training data. Furthermore, MLIPs often lack sufficient generalization capability and robustness during MD simulations, i.e., they are sensitive to outliers and local perturbations of atomic structures. This sensitivity of ML models is caused by existing data sets and data generation techniques not providing sufficient coverage of configurational and compositional spaces.

\begin{figure*}[t]
    \centering
    \includegraphics[width=0.95\textwidth]{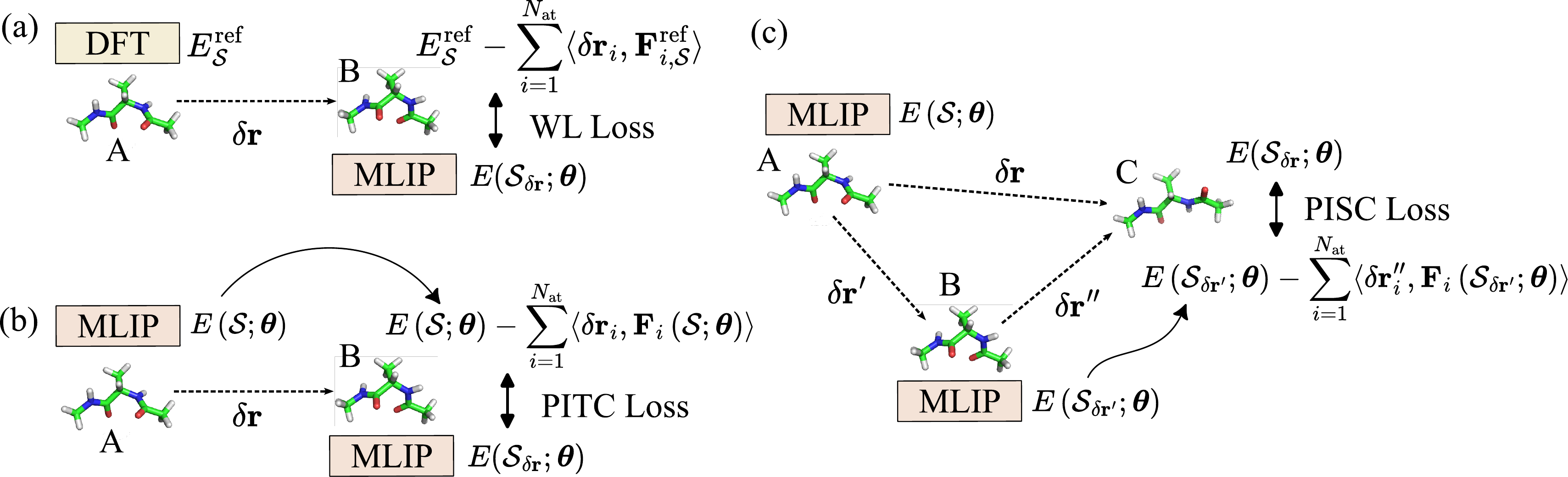}
    \caption{\textbf{Schematic illustration of physics-informed weakly supervised losses used in this work.} (a) Taylor-expansion-based weak label (WL) loss with approximate labels obtained from reference energies and atomic forces~\citep{cooper2020efficient}. (b) Physics-inspired Taylor-expansion-consistency (PITC) loss with approximate labels obtained from energies and atomic forces predicted by an MLIP. (c) Physics-inspired spatial consistency (PISC) loss with approximate labels obtained from energies and atomic forces predicted by an MLIP. Here, $E(\mathcal{S}; \boldsymbol{\theta})$ and $\mathbf{F}_i(\mathcal{S}; \boldsymbol{\theta})$ denote the potential energy and atomic forces predicted by an MLIP parametrized by $\boldsymbol{\theta}$, $\mathcal{S}$ and $\mathcal{S}_{\delta\mathbf{r}}$ define the original atomic structure and the one perturbed by $\delta\mathbf{r}$.}
    \label{fig:weakly-supervised}
\end{figure*}

\textbf{Contributions.} This paper addresses these challenges using a physics-informed weakly supervised learning (PIWSL) approach. Our method is designed to learn an MLIP, which can accurately predict the potential energy and atomistic forces for an atomic system exposed to local perturbations. In particular, our contributions are as follows: (i) We introduce PIWSL based on basic physical principles, such as the concept of conservative forces. We combine it with extrapolating the potential energy via a Taylor expansion and derive two novel physics-informed loss functions, schematically illustrated in \figref{fig:weakly-supervised} (b) and (c). Particularly, we obtain physics-informed Taylor-expansion-based consistency (PITC) and physics-informed spatial consistency (PISC) losses, which build the basis for the PIWSL approach. (ii) By conducting extensive experiments, we demonstrate that PIWSL facilitates the training of MLIPs without access to large training data sets. Furthermore, we show that MLIPs trained with PIWSL are more robust during MD simulations compared to those trained without it. (iii) We also observe that PIWSL improves accuracy in predicted total energies and atomic forces, even without access to force labels. This scenario is particularly relevant for fine-tuning foundation MLIP models with reference ab initio methods where atomic force calculations are computationally intractable or entirely unavailable~\citep{Smith2019, smith2020ani, Zaverkin2023}. Thus, our results open new possibilities for training MLIPs using highly accurate energy labels, such as those obtained by extrapolating CCSD(T) energies to the complete basis set (CBS) limit~\citep{Hobza2002, Feller2006}. (iv) Finally, PIWSL mitigates sensitivity issues during MD simulations associated with limited sizes of available data sets by taking into account the potential energy response to local perturbations in atomic structures.

\section{Related Work \label{sec:related_work}}

\textbf{Machine-Learned Interatomic Potentials.} There is a growing interest in using ML-based models for investigating molecular and material systems as they allow performing atomistic simulations with an accuracy on par with first-principles methods but at a fraction of the computational cost. The field of machine-learned interatomic potentials (MLIPs) emerged over two decades ago~\citep{blank1995neural} and has been one of the most active research directions since then~\citep{behler2007generalized, artrith2011high, ARTRITH2016135, Smith2017, Shapeev2016, schutt2017schnet, thomas2018tensor, Unke2019, drautz2019atomic, Zaverkin2020, Zaverkin2021b, thomas2018tensor, schutt2021equivariant, shuaibi2021rotation, escn, equiformer_v2, batzner2023, musaelian2023learning, batatia2022mace, Zaverkin2024b}. The development of local higher-body-order representations~\citep{Shapeev2016, drautz2019atomic, Zaverkin2020, Zaverkin2021b} and the emergence of equivariant message-passing neural networks (MPNNs)~\citep{thomas2018tensor, schutt2021equivariant, shuaibi2021rotation, escn, equiformer_v2, batzner2023, musaelian2023learning, batatia2022mace, Zaverkin2024b} significantly advanced the field. These methods enable the cost-efficient generation of accurate MLIPs for modeling interactions in many-body atomic systems and account for crucial inductive biases as the invariance of the potential energy under rotation.

\textbf{Physics-Informed Machine Learning.} Physics-informed ML aims to model physical systems using data-driven techniques and incorporates physics principles into ML-based models. For example, MLIPs based on equivariant MPNNs enforce the invariance of the potential energy under rotation and use equivariant features to enrich the building of many-body contributions to it~\citep{thomas2018tensor, batzner2023, batatia2022mace, musaelian2023learning, equiformer_v2, Zaverkin2024b}. Furthermore, physics constraints can be integrated via auxiliary loss functions, prompting ML models to learn important physical relationships, as demonstrated for physics-informed neural networks (PINNs)~\citep{2019JCoPh.378..686R, cai2022physics},  which learn to model solutions of partial differential equations by minimizing residuals during training. Applying physics-informed ML to molecular modeling has gained attraction in both ML and computational chemistry communities~\citep{godwin2021simple, ni2023sliced}. As such, prior work~\citep{cooper2020efficient} has motivated our current research and is discussed in more detail in subsequent sections.

\section{Background and Problem Definition \label{sec:background-and-problem}}

\textbf{Machine-Learned Interatomic Potentials.} An atomic configuration, denoted as $\mathcal{S} = \{\mathbf{r}_i, Z_i\}_{i=1}^{N_\mathrm{at}}$, contains $N_\mathrm{at}$ atoms and is defined by atom positions $\mathbf{r}_i \in \mathbb{R}^3$ and atom types $Z_i \in \mathbb{N}$. We consider mapping atomic configurations to scalar energies, i.e., $f_{\boldsymbol{\theta}}: \mathcal{S} \mapsto E \in \mathbb{R}$ with $\boldsymbol{\theta}$ denoting trainable parameters. We define $E(\mathcal{S}; \boldsymbol{\theta})$ as the energy predicted by an MLIP for an atomic configuration $\mathcal{S}$. For most MLIPs, atomic forces are computed as the negative gradients of the potential energy with respect to atom positions, i.e., $\mathbf{F}_i\left(\mathcal{S}; \boldsymbol{\theta}\right) = -\nabla_{\mathbf{r}_i} E\left(\mathcal{S}; \boldsymbol{\theta}\right)$. In this way, these MLIPs ensure that the resulting forces are conservative (curl-free) and the total energy is conserved during a dynamic simulation. However, some models are designed to predict atomic forces directly~\citep{hu2021forcenet,escn,equiformer_v2,ocp_dataset}. While this approach avoids expensive gradient computations, it violates the law of energy conservation~\citep{chmiela2017machine}.

Trainable parameters $\boldsymbol{\theta}$ are optimized by minimizing loss functions on training data $\mathcal{D}$ comprising a total of $N_\mathrm{train}$ atomic configurations $\{\mathcal{S}^{(k)}\}_{k=1}^{N_\mathrm{train}}$ as well as their energies $\{ E^\mathrm{ref}_{\mathcal{S}}\}_{\mathcal{S} \in \mathcal{D}}$ and atomic forces $\{\{\mathbf{F}_{i,\mathcal{S}}^\mathrm{ref}\}_{i=1}^{N_\mathrm{at}}\}_{\mathcal{S} \in \mathcal{D}}$
\begin{align}
    \label{eq:MLIP-loss}
    &\mathcal{L}\left(\mathcal{D};\boldsymbol{\theta}\right) = \sum_{\mathcal{S} \in \mathcal{D}} L\left(S;\boldsymbol{\theta}\right) 
    \nonumber
    \\
    &= \sum_{\mathcal{S} \in \mathcal{D}} \Big[C_\mathrm{e} \ell\left(E\left(\mathcal{S}; \boldsymbol{\theta}\right), E^\mathrm{ref}_\mathcal{S}\right) + C_\mathrm{f} \sum_{i=1}^{N_\mathrm{at}} \ell\left(\mathbf{F}_i (\mathcal{S}; \boldsymbol{\theta}), \mathbf{F}_{i,\mathcal{S}}^\mathrm{ref} \right) \Big].
\end{align}
Here, $\ell$ denotes a point-wise loss function such as the absolute and squared error between the predicted and reference total energies and atomic forces. Typically, reference energies $E^\mathrm{ref}_{\mathcal{S}}$ and atomic forces $\mathbf{F}_{i,\mathcal{S}}^\mathrm{ref}$ are provided by ab initio or first-principles methods such as CC or DFT, respectively. The relative contributions of energies and forces in \eqref{eq:MLIP-loss} are balanced with the coefficients $C_\mathrm{e}$ and $C_\mathrm{f}$.

\textbf{Weakly Supervised Learning.} Generating many reference labels with a first-principles approach is challenging due to the high computational cost. Furthermore, the calculation of atomic forces can be infeasible for some high-accuracy ab initio methods, e.g., for CCSD(T)/CBS. In this work, we focus on weakly supervised learning methods to improve the performance of MLIPs in scenarios when only a limited amount of data is available. These involve the generation of approximate but physically motivated total energies for atomic structures generated by small perturbations of their atomic positions, i.e., $\mathcal{S}_{\dr} = \{\mathbf{r}_i + \dr_i, Z_i\}_{i=1}^{N_\mathrm{at}}$ with a perturbation vector $\dr$, where $\dr_i$ is the perturbation vector for atom $i$. Approximate labels are computed with MLIPs during their training.

\section{Physics-informed Weakly Supervised Learning \label{sec:physics-informed-wsl}}

For MLIPs, the generation of approximate labels employed in weakly supervised losses is highly non-trivial. Small perturbations in atomic structures can significantly change energies and atomic forces. Thus, standard approaches that are effective for many ML tasks~\citep{yang2022survey} are typically not applicable to MLIPs. To address this problem, we propose a physics-informed weakly supervised learning approach that involves (i) a Taylor expansion of the potential energy for computing the response to atomic perturbations and (ii) spatial consistency to estimate the displaced potential energy based on the concept of conservative forces. We finally introduce the PIWSL loss term, combining both classes of weakly supervised loss functions with the supervised loss.

\subsection{Physics-Informed Taylor-Expansion-Based Consistency Loss \label{sec:taylor-loss}}

This section introduces the physics-informed Taylor-expansion-based consistency (PITC) loss. Particularly, we relate the energy predicted directly for a displaced atomic configuration with the energy obtained by the Taylor expansion from the original configuration; see \figref{fig:weakly-supervised} (b). We estimate the energy for an atomic structure $\mathcal{S}$ drawn from the training data set with atomic positions displaced by a vector $\dr$: $\mathcal{S}_{\dr} = \{\mathbf{r}_i + \dr_i, Z_i\}_{i=1}^{N_\mathrm{at}}$. For this atomic configuration, we expand the energy predicted by an MLIP in its second-order Taylor series around the atomic perturbation vector $\dr_i$ and obtain
\begin{align}
    \label{eq:taylor-wsl}
        &E\left(\mathcal{S}_{\dr}; \boldsymbol{\theta}\right) \approx  E\left(\mathcal{S}; \boldsymbol{\theta}\right) 
        - \sum_{i=1}^{N_\mathrm{at}} \left[ \langle \dr_i, \mathbf{F}_i \left(\mathcal{S}; \boldsymbol{\theta}\right) \rangle \right.
        \nonumber
        \\
        & \left. + k_{\rm 2nd} \langle \dr_i, \mathbf{F}_i\left(\mathcal{S}_{\dr}; \boldsymbol{\theta}\right) - \mathbf{F}_i\left(\mathcal{S}; \boldsymbol{\theta}\right) \rangle \right] + \mathcal{O}\left(\|\dr\|^3\right),
\end{align}
where $\langle \cdot \rangle$ denotes the inner product. Here, we used atomic forces, defined as the negative gradients of the potential energy. The parameter $k_{\rm 2nd}$ controls the contribution of the second-order term. Setting $k_{\rm 2nd} = 1/2$ recovers the exact second-order Taylor expansion, while $k_{\rm 2nd} = 0$ leads to the first-order approximation\footnote{A more detailed derivation is provided in \secref{sec:PISC-variance}.}. Using approximate labels $E\left(\mathcal{S}_{\dr}; \boldsymbol{\theta}\right)$, we define the PITC loss as
\begin{align}
    \label{eq:taylor-wsl-loss}
    &L_\mathrm{PITC}\left(\mathcal{S};\boldsymbol{\theta}\right)  =  \ell\Big( E\left(\mathcal{S}_{\dr}; \boldsymbol{\theta}\right), E\left(\mathcal{S}; \boldsymbol{\theta}\right) 
    \nonumber
    \\
    & - \sum_{i=1}^{N_\mathrm{at}} \langle \dr_i, (1 - k_{\rm 2nd}) \mathbf{F}_i \left(\mathcal{S}; \boldsymbol{\theta}\right) + k_{\rm 2nd} \mathbf{F}_i\left(\mathcal{S}_{\dr}; \boldsymbol{\theta}\right) \rangle \Big),
\end{align}
where $\ell$ denotes a point-wise loss for regression problems and $\delta \boldsymbol{r}$ is randomly-sampled or determined adversarially; see \secref{sec:sub-perturbation-types} for more details. Hence, whenever we encounter a structure $\mathcal{S}$ in a batch during training, a new $\delta \boldsymbol{r}$ is computed for each $\mathcal{S}$. Empirically, second-order terms become important when MLIP prediction errors reach the first-order accuracy level, meaning that first-order terms alone are no longer sufficient to improve accuracy or when the data primarily consists of relaxed structures.

\subsection{Physics-Informed Spatial-Consistency Loss \label{sec:conservative-force}}

This section introduces a physics-informed approach for generating weak labels based on the concept of conservative forces. Thus, we leverage the fact that the energy difference between two points on the potential energy surface is independent of the path taken between them. We consider two paths from a reference point to the same target point, composed of three perturbation vectors in total. We estimate the potential energy at the target point via \eqref{eq:taylor-wsl}. An example of two paths is demonstrated in \figref{fig:weakly-supervised} (c). The figure relates the energy obtained when displacing atomic positions of the original configuration $\mathcal{S}$ (denoted by A in the figure) by $\dr$ (from configuration A to C) with the energy obtained through consecutive perturbations $\dr^\prime$ (from configuration A to B) and $\dr^{\prime\prime}$ (from configuration B to C).

For the first path, we directly predict the energy with an MLIP, i.e., $E\left(\mathcal{S}_{\dr}; \boldsymbol{\theta}\right)$, which is related to the approximated energy at $\mathbf{r} + \dr$ using \eqref{eq:taylor-wsl-loss} through PITC loss. For the second path, we directly compute the energy $E\left(\mathcal{S}_{\dr^\prime}; \boldsymbol{\theta}\right)$ for atomic positions displaced by $\dr^\prime$ and use it to approximate $E\left(\mathcal{S}_{\dr}; \boldsymbol{\theta}\right)$ after applying the second perturbation vector $\dr^{\prime\prime} \equiv \dr - \dr^{\prime}$. The physics-informed spatial consistency (PISC) loss can be defined as
\begin{align}
    \label{eq:spatial-consis-loss}
    L_\mathrm{PISC}\left(\mathcal{S};\boldsymbol{\theta}\right) = \ell\Big( E\left(\mathcal{S}_{\dr}; \boldsymbol{\theta}\right), E_{\rm PITC}\left(\mathcal{S}_{\dr^\prime}, \dr^{\prime\prime}; \boldsymbol{\theta}\right) \Big),
\end{align}
where $E_{\rm PITC}\left(\mathcal{S}_{\dr}, \dr^{\prime}; \boldsymbol{\theta}\right)$ is the potential energy estimated via PITC formula in \eqref{eq:taylor-wsl} from the configuration $\mathcal{S}_{\dr}$ perturbed by $\dr^{\prime}$. After joint training of PITC and PISC losses, the three different estimations at $\mathcal{S}_{\dr}$ become spatially consistent. Note that our conservative forces-based approach is not limited to relations between two perturbation paths or three perturbation vectors. We discuss several other possible configurations in \secref{sec:other-consistency}.

\subsection{Combined Physics-Informed Weakly Supervised Loss}

Together with the usual MLIP loss function defined in \eqref{eq:MLIP-loss}, the overall objective, which we refer to as the PIWSL loss, can be written as
\begin{align}
    \label{eq:PILWS-loss}
    &\argmin_{\boldsymbol{\theta}}\tilde{\mathcal{L}}\left(\mathcal{D}; \boldsymbol{\theta} \right) 
    = \argmin_{\boldsymbol{\theta}} \sum_{\mathcal{S} \in \mathcal{D}} \left(L\left(\mathcal{S}; \boldsymbol{\theta}\right) \right.
    \nonumber
    \\
    & \left. + C_{\mathrm{PITC}} L_{\mathrm{PITC}}\left(\mathcal{S}; \boldsymbol{\theta}\right) + C_{\mathrm{PISC}} L_{\mathrm{PISC}} \left(\mathcal{S}; \boldsymbol{\theta}\right)\right),
\end{align}
where $C_{\rm PITC}$ and $C_{\rm PISC}$ are the weights of the weakly supervised PITC and PISC losses.

\subsection{Determining Perturbation Directions and Magnitudes \label{sec:sub-perturbation-types}}

The effectiveness of the proposed approach depends on appropriate choices of the perturbation vectors $\dr$. We introduce and justify various strategies for generating the perturbations used in \eqref{eq:taylor-wsl-loss} and \eqref{eq:spatial-consis-loss}. Any vector $\dr$ can be written as $\dr \equiv \epsilon \mathbf{g}/\|\mathbf{g}\|_2$, where $\epsilon$ is the magnitude of $\dr$ and $\mathbf{g}/\|\mathbf{g}\|_2$ is its direction. Physical constraints can limit $\epsilon$. Specifically, we can obtain the maximum perturbation length from the validity of the Taylor expansion in \eqref{eq:taylor-wsl}, which, as discussed in \secref{sec:vis-PIWSL}, is typically given as at most 30\% of the original bond length whose shortest example is the bond between carbon and hydrogen atoms, about 1.09 \AA; see also \figref{fig:enter-label} (c) and (d). The specific values of $\epsilon$ chosen for our experiments are provided in \secref{sec:setup}.

To determine $\mathbf{g}/\|\mathbf{g}\|_2$ we explore two strategies. First, we compute it as the unit vector of a perturbation vector sampled from the uniform distribution on the interval $(-1, 1)$ for each direction
\begin{equation}
    \label{eq:adv-random}
    \dr_{\rm rnd} \equiv \epsilon \mathbf{g}/\|\mathbf{g}\|_2.
\end{equation}
Second, we compute an adversarial direction, as proposed by~\citet{goodfellow2014explaining,miyato2018virtual}, which involves defining it as the direction (the gradients) in which the loss error increases the most at the current atom coordinates $\mathbf{r}$ and for the current predicted energy. Assuming the norm of adversarial perturbation as $L_2$, the adversarial direction can be approximated by \citep{miyato2018virtual}
\begin{equation}
    \delta\mathbf{r}_{\rm adv} \equiv \epsilon \mathbf{g}/\|\mathbf{g}\|_2, \ {\rm where} \ \mathbf{g} = \nabla_{\mathbf{r}} L_{\rm dist}(\mathbf{y}^{\rm pred}, \mathbf{y}^{\rm ref}), 
    \label{eq:adv-direc}
\end{equation}
where $L_{\rm dist}$ is a distance measure function to be maximized by adding $\dr_{\rm adv}$, with $\mathbf{y}^{\rm pred}$ and $\mathbf{y}^{\rm ref}$ being the ML model prediction and the reference values. Due to their computational efficiency, we mainly use \eqref{eq:adv-random} in our experiments. A quantitative comparison between the random and adversarial directions is provided in \secref{sec:analysis}.

\section{Experiments \label{sec:experiment}}

We evaluate our method through extensive experiments with the following objectives: (1) comparing PIWSL with existing baselines, (2) analyzing the impact of the PIWSL using the aspirin molecule, including MD simulations, (3) assessing PIWSL's ability to improve foundation model finetuning on a sparse dataset, particularly when energy and force predictions when force labels are inaccessible.\footnote{In \secref{sec:analysis}, we provide additional results on: (4) a comparison with a prior weakly supervised approach, (5) an ablation study, and (6) a comparison between random and adversarial perturbation vectors.} Our experiments mainly use the first-order PITC except for the MD17 and MD22 data sets, where target accuracy requires considering the second-order term.\footnote{The effect of the second order term is analyzed in \secref{sec:PIWSL-order-comparison}.}

\begin{table*}[t!]
    \caption{\textbf{Energy (E) and force (F) root-mean-square errors (RMSEs) for the ANI-1x data set.} The results are obtained by averaging over three independent runs. Energy RMSE is given in kcal/mol, while force RMSE is in kcal/mol/\AA. \label{tab:ani-1x-results}}
    \begin{center}
    \resizebox{0.8\textwidth}{!}{
    \begin{tabular}{lcrrrrrr}
    \toprule
                                    &   & \multicolumn{3}{c}{$N_\mathrm{train} = 100$}                                   &  \multicolumn{3}{c}{$N_\mathrm{train} = 1000$}                                            \\
                                    &   & Baseline           & Noisy Nodes               & PIWSL                         & Baseline                  & Noisy Nodes                   & PIWSL                         \\
    \cmidrule(lr){1-2} \cmidrule(lr){3-5} \cmidrule(lr){6-8}
    \multirow[c]{2}{*}{SchNet}      & E & 65.09 $\pm$ 2.42   & {\bf 57.39 $\pm$ 0.05}    & 60.30 $\pm$ 1.77              & 31.49 $\pm$ 0.01          & {\bf 31.10 $\pm$ 0.00}        & 31.50 $\pm$ 0.00              \\
                                    & F & 29.06 $\pm$ 0.19   & {\bf 25.62 $\pm$ 0.01}    & 28.20 $\pm$ 0.60              & 18.94 $\pm$ 0.01          & {\bf 18.10 $\pm$ 0.00}        & 18.93 $\pm$ 0.00              \\
    \cmidrule(lr){1-2} \cmidrule(lr){3-5} \cmidrule(lr){6-8}
    \multirow[c]{2}{*}{PaiNN}       & E & 168.01 $\pm$ 1.22  & 464.55 $\pm$ 6.91         & {\bf 109.89 $\pm$ 11.46}      & 56.62 $\pm$ 2.80          & 305.76 $\pm$ 33.93            & {\bf 24.53 $\pm$ 0.48}        \\  
                                    & F & 21.33 $\pm$ 0.10   & 20.82 $\pm$ 0.03          & {\bf 18.76 $\pm$ 0.30}        & 12.96 $\pm$ 0.06          & 14.25 $\pm$ 0.18              & {\bf 11.43 $\pm$ 0.05}        \\
    \cmidrule(lr){1-2} \cmidrule(lr){3-5} \cmidrule(lr){6-8}
    \multirow[c]{2}{*}{SpinConv}    & E & 162.14 $\pm$ 7.55  & 147.73 $\pm$ 2.23         & {\bf 130.97 $\pm$ 8.58}       & 43.59 $\pm$ 1.71          & 299.33 $\pm$ 419.10           & {\bf 39.44 $\pm$ 1.31}        \\
                                    & F & 21.22 $\pm$ 0.43   & {\bf 21.08 $\pm$ 0.43}    & 21.61 $\pm$ 0.44              & 14.51 $\pm$ 1.07          & 15.83 $\pm$ 0.75              & {\bf 13.59 $\pm$ 0.20}        \\
    \cmidrule(lr){1-2} \cmidrule(lr){3-5} \cmidrule(lr){6-8}
    \multirow[c]{2}{*}{eSCN}        & E & 214.52 $\pm$ 7.55  & 521.92 $\pm$ 12.05        & {\bf 183.70 $\pm$ 9.79}       & 59.59 $\pm$ 8.92          & 241.34 $\pm$ 20.16            & {\bf 21.03 $\pm$ 0.56}        \\
                                    & F & 20.07 $\pm$ 0.27   & 23.68 $\pm$ 0.11          & {\bf 19.69  $\pm$ 0.05}       & 12.50 $\pm$ 0.78          & 14.42 $\pm$ 0.84              & {\bf 11.83 $\pm$ 0.12}        \\
    \cmidrule(lr){1-2} \cmidrule(lr){3-5} \cmidrule(lr){6-8}
    \multirow[c]{2}{*}{Equiformer}  & E & 398.71 $\pm$ 13.69 & 632.38 $\pm$ 0.11         & {\bf 154.98 $\pm$ 8.83}       & 54.52 $\pm$ 4.52          & 854.33 $\pm$ 317.7            & {\bf 20.89 $\pm$ 0.50}        \\
                                    & F & 20.71 $\pm$ 0.05   & 21.82 $\pm$ 0.01          & {\bf 20.55 $\pm$ 0.05}        & 10.10 $\pm$ 0.00          & 24.79 $\pm$ 2.05              & {\bf 9.68 $\pm$ 0.03}         \\
    \bottomrule
    \end{tabular}}
    \end{center}
\end{table*}

\subsection{Models and Data Sets}

We trained the following representative models that are provided in the Open Catalyst code base \citep{ocp_dataset}: SchNet ~\citep{schutt2017schnet}, PaiNN~\citep{schutt2021equivariant}, SpinConv~\citep{shuaibi2021rotation}, eSCN~\citep{escn}, and Equiformer v2~\citep{equiformer_v2}, covering MLIPs with a smaller  (SchNet, SpinConv, PaiNN) and larger number of parameters (eSCN, Equiformer v2). Moreover, we also considered the MACE model \citep{batatia2022mace}, a state-of-the-art model that we use to evaluate the impact of PIWSL on the MD17(CCSD) and MD22 data set. Unless otherwise mentioned and except for SchNet, forces are directly predicted and not computed through the negative gradient of the energy. The results where forces are computed as negative energy gradients are analyzed in \secref{sec:no-force-label} and \secref{sec:GF-experiments}.

To evaluate the effect and dependency of the physics-informed weakly supervised approach in detail, we performed the training on various data sets: ANI-1x as a heterogeneous molecular data set~\citep{smith2020ani}, TiO$_2$ as a data set for inorganic materials~\citep{ARTRITH2016135}, the revised MD17 (rMD17) data set containing small molecules with sampled configurational spaces for each~\citep{chmiela2017machine,chmiela2018towards,Christensen2020}, the MD22 data set containing larger molecules \cite{md22}, and LMNTO as another material data set \citep{cooper2020efficient}; the benchmark results for rMD17, MD22, and LMNTO are provided in \secref{sec:add-benchmark}. The detailed description of each data set is provided in \secref{sec:dataset-explanation}.

\subsection{Benchmark Results \label{sec:benchmark}}

We compare models trained using the PIWSL loss (see \eqref{eq:PILWS-loss}) with baseline models trained using the standard supervised loss only (see \eqref{eq:MLIP-loss}). We also compare our approach to a recently proposed data augmentation method that incorporates the task of denoising random perturbations of the atomic coordinates into the learning objective (NoisyNode)~\citep{godwin2021simple}. 
A comparison with weak label method \citep{cooper2020efficient} is provided in \secref{sec:analysis}.
More details on the setup are provided in \secref{sec:setup}. In the following, all evaluation metrics are computed for the test data set.

\textbf{Heterogeneous Molecular Data Set -- ANI-1x.} The results provided in \tabref{tab:ani-1x-results} show that our approach improves the baseline models' performance in almost all cases. In particular, the error reduction for the predicted energies is often between $10$~\% and more than $50$~\%. Interestingly, we observe an improved accuracy for potential energies and atomic forces because we include force prediction in PITC and PISC losses, different from the previous work \citep{cooper2020efficient}. In most cases, except for SchNet, the data augmentation method (NoisyNode) deteriorates the accuracy of the MLIPs because it does not incorporate the proper response of the energy and atomic forces to the perturbation of atomic positions.

\begin{table*}[t!]
\caption{\textbf{Energy (F) and force (F) root-mean-square errors (RMSEs) for the TiO$_2$ data set.} The results are obtained by averaging over three independent runs. Energy RMSE is given in kcal/mol, while force RMSE is in kcal/mol/\AA. \label{tab:tio2-results}}
    \begin{center}
    \resizebox{0.8 \textwidth}{!}{
    \begin{tabular}{lcrrrrrr}
    \toprule
                                    &   & \multicolumn{3}{c}{$N_\mathrm{train} = 100$}                                        &  \multicolumn{3}{c}{$N_\mathrm{train} = 2000$}  \\
                                    &   & Baseline          & Noisy Nodes                    & PIWSL                          & Baseline                                    & Noisy Nodes                    & PIWSL                         \\
    \cmidrule(lr){1-2} \cmidrule(lr){3-5} \cmidrule(lr){6-8}
    \multirow[c]{2}{*}{SchNet\textsuperscript{\emph{a}}}      & E & 17.21 $\pm$ 0.00  & 19.68 $\pm$ 0.00  & {\bf 17.08 $\pm$ 0.00}               & 10.16 $\pm$ 0.00 & 44.44 $\pm$ 0.00              & {\bf 10.14 $\pm$ 0.00}        \\
                                    & F & 2.84 $\pm$ 0.00   & {\bf 2.70 $\pm$ 0.00}          & 2.83 $\pm$ 0.00                & 1.87 $\pm$ 0.00   & 7.45 $\pm$ 0.00               & {\bf 1.85 $\pm$ 0.00}         \\
    \cmidrule(lr){1-2} \cmidrule(lr){3-5} \cmidrule(lr){6-8}
    \multirow[c]{2}{*}{PaiNN\textsuperscript{\emph{b}} }       & E & 14.41 $\pm$ 0.16  & n/a\textsuperscript{\emph{b}}  & {\bf 13.95 $\pm$ 0.09} & 2.44 $\pm$ 0.03 & n/a\textsuperscript{\emph{b}}  & {\bf 2.30 $\pm$ 0.10} \\
                                    & F & 1.59 $\pm$ 0.01 & n/a\textsuperscript{\emph{b}}  & {\bf 1.56 $\pm$ 0.01} & 0.27 $\pm$ 0.02   & n/a\textsuperscript{\emph{b}}  & {\bf 0.24 $\pm$ 0.00} \\
    \cmidrule(lr){1-2} \cmidrule(lr){3-5} \cmidrule(lr){6-8}
    \multirow[c]{2}{*}{SpinConv}    & E & 20.00 $\pm$ 0.42  & 18.76 $\pm$ 0.74               & {\bf 16.98 $\pm$ 0.99}         & 2.78 $\pm$ 0.67                             & 2.76 $\pm$ 0.42                & {\bf 2.05 $\pm$ 0.39}         \\
                                    & F & 1.58 $\pm$ 0.03   & {\bf 1.53 $\pm$ 0.03}          & 1.59 $\pm$ 0.03                & 0.67 $\pm$ 0.15                             & {\bf 0.61 $\pm$ 0.09}                & {\bf 0.61 $\pm$ 0.05}         \\
    \cmidrule(lr){1-2} \cmidrule(lr){3-5} \cmidrule(lr){6-8}
    \multirow[c]{2}{*}{eSCN}        & E & 16.41 $\pm$ 1.10  & 20.92 $\pm$ 0.00               & {\bf 12.63 $\pm$ 0.78}         & 1.78 $\pm$ 0.07                             & 20.90 $\pm$ 0.00               & {\bf 0.90 $\pm$ 0.09}         \\
                                    & F & 1.57 $\pm$ 0.04   & 1.66 $\pm$ 0.00                & {\bf 1.44 $\pm$ 0.03}          & 0.42 $\pm$ 0.17                             & 1.66 $\pm$ 0.00                & {\bf 0.16 $\pm$ 0.01}         \\
    \cmidrule(lr){1-2} \cmidrule(lr){3-5} \cmidrule(lr){6-8}
    \multirow[c]{2}{*}{Equiformer}  & E & 18.21 $\pm$ 0.02  & 19.06 $\pm$ 0.02               & {\bf 13.93 $\pm$ 0.09}         & 1.73 $\pm$ 0.05                             & 18.54 $\pm$ 0.10               & {\bf 1.27 $\pm$ 0.02}         \\
                                    & F & 1.56 $\pm$ 0.01   & 1.64 $\pm$ 0.00                & {\bf 1.51 $\pm$ 0.19}          & {\bf 0.13 $\pm$ 0.01}                       & 1.59 $\pm$ 0.00                & {\bf 0.13 $\pm$ 0.00}         \\
    \bottomrule
    \end{tabular}}
    \end{center}
    \footnotesize{\textsuperscript{\emph{a}} We used a larger batch size of $32$ for SchNet since we obtained extremely high errors for the batch size of $4$. A more detailed discussion of the experimental results for SchNet is provided in \secref{sec:add-benchmark}. \\
    \textsuperscript{\emph{b}} Because of a numerical instability of PaiNN when perturbing atomic coordinates, the cutoff radius is reduced from 12~\AA{} to 5~\AA{} in this experiment. Predicted values become n/a when atomic configurations are perturbed.}
\end{table*}

\begin{figure*}[t!]
    \centering    
    \includegraphics[width=\textwidth]{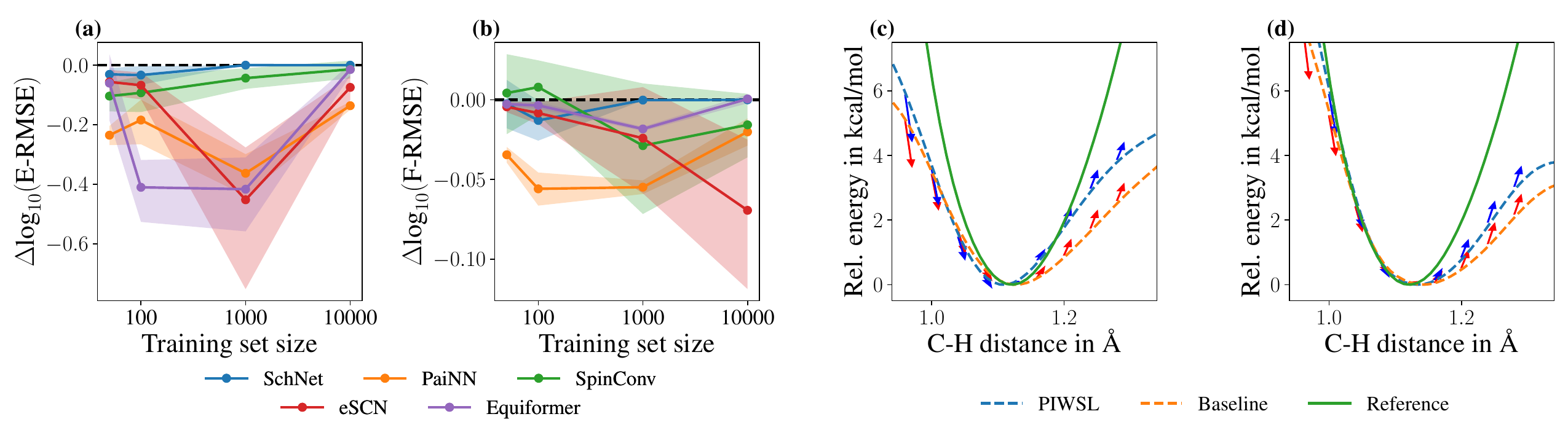}
    \caption{\textbf{(a, b) Relative performance gains for MLIPs trained with PIWSL compared to those trained without it and (c, d) potential energy profiles for a C--H bond of the aspirin molecule.} Relative performance gains are evaluated for (a) energy (E-) and (b) force (F-) RMSEs. These results are presented for the ANI-1x data set. Potential energy profiles for a C--H bond of the aspirin molecule are presented for models trained using (c) 100 and (d) 200 configurations. The red and blue arrows indicate the direction from the original structure ($E(\mathcal{S};\boldsymbol{\theta})$) to the perturbed one ($E(\mathcal{S}_{\dr};\boldsymbol{\theta})$), as defined by \eqref{eq:taylor-wsl}, for the baseline and PIWSL model predictions, respectively.}
    \label{fig:enter-label}
\end{figure*}

\textbf{Training Data Set Size Dependence -- ANI-1x.} We train MLIPs with training set sizes of $[50, 10^2, 10^3, 10^4, 10^5, $ $10^6, 5 \times 10^6]$. The results for training data sizes of $10^5$, $10^6$, and $5 \times 10^6$ are provided in \secref{sec:add-benchmark}. The results are plotted in \figref{fig:enter-label} (a) and (b). Although the observed error reduction depends strongly on the type of MLIP used, the benefit of the weakly supervised losses often decreases slightly with the number of training samples. This result can be expected as the area covered by the weakly supervised losses is also gradually covered by the reference data as the number of training samples increases. Moreover, the gain in accuracy of energy predictions is more significant than that for forces trained only indirectly through the consistency constraint in PITC; see \eqref{eq:taylor-wsl-loss}. Finally, it is shown that the improvement is more significant for highly parameterized MLIPs, which benefit the most from increasing the training data size through PIWSL.

\textbf{Inorganic Bulk Materials -- TiO$_2$.} Titanium dioxide (TiO$_2$) is a highly relevant metal oxide for industrial applications, featuring several high-pressure phases. Thus, ML models should be able to predict total energies and atomic forces for various high-pressure phases of TiO$_2$, considering periodic boundaries (relevant when aggregating over the local atomic neighborhood). The results for trained models are provided in \tabref{tab:tio2-results}. Similar to the ANI-1x data set, our approach improves the accuracy of predicted energies and atomic forces. Interestingly, even when the error in the potential energy for 2000 training configurations reaches small RMSE values, close to or even less than 1~kcal/mol in predicted energy, the PIWSL still reduces the error. This observation indicates strong evidence of the effectiveness of PIWSL applied to bulk materials.

\subsection{Qualitative Impact of PIWSL \label{sec:vis-PIWSL}}

\begin{figure*}[t!]
    \centering
    \includegraphics[width=0.6\textwidth]{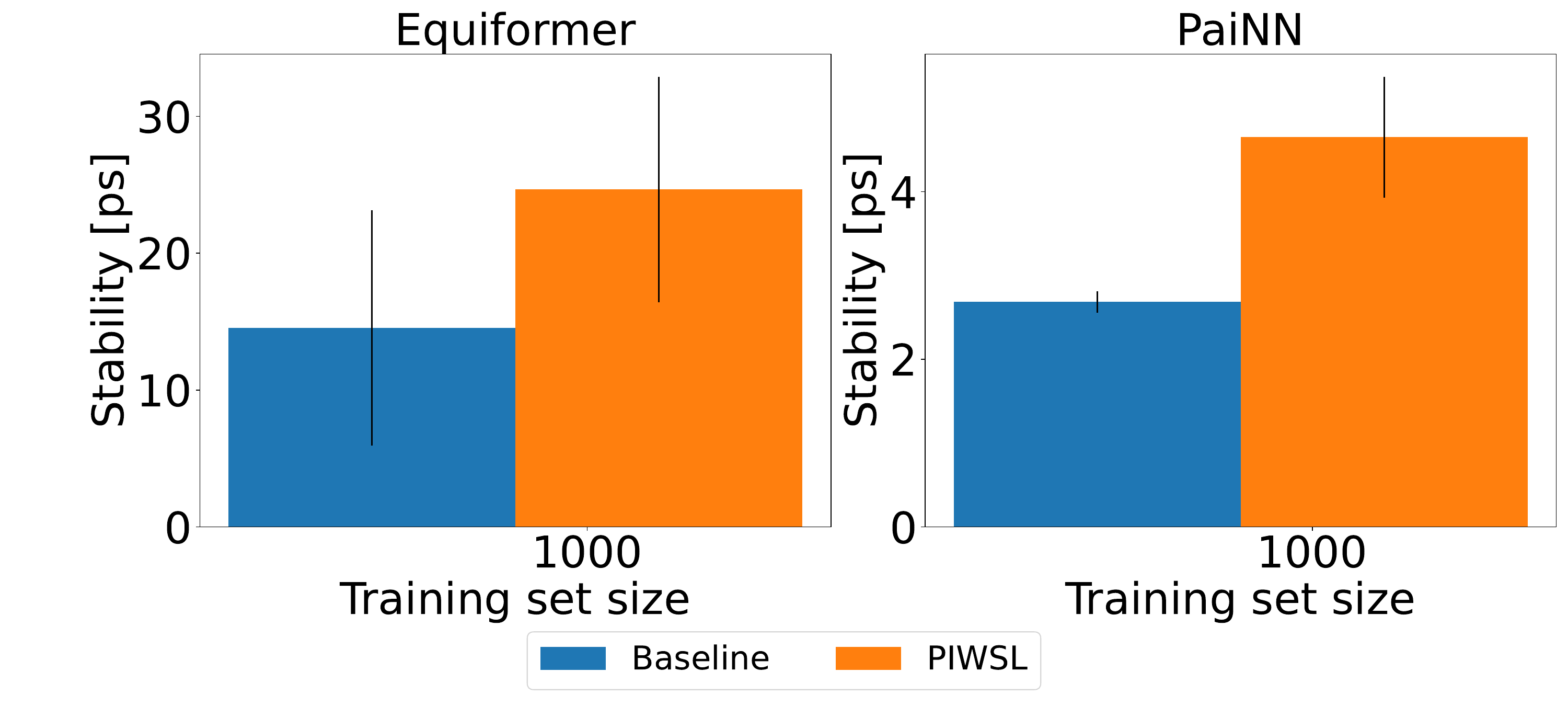}
    \includegraphics[width=0.3\textwidth]{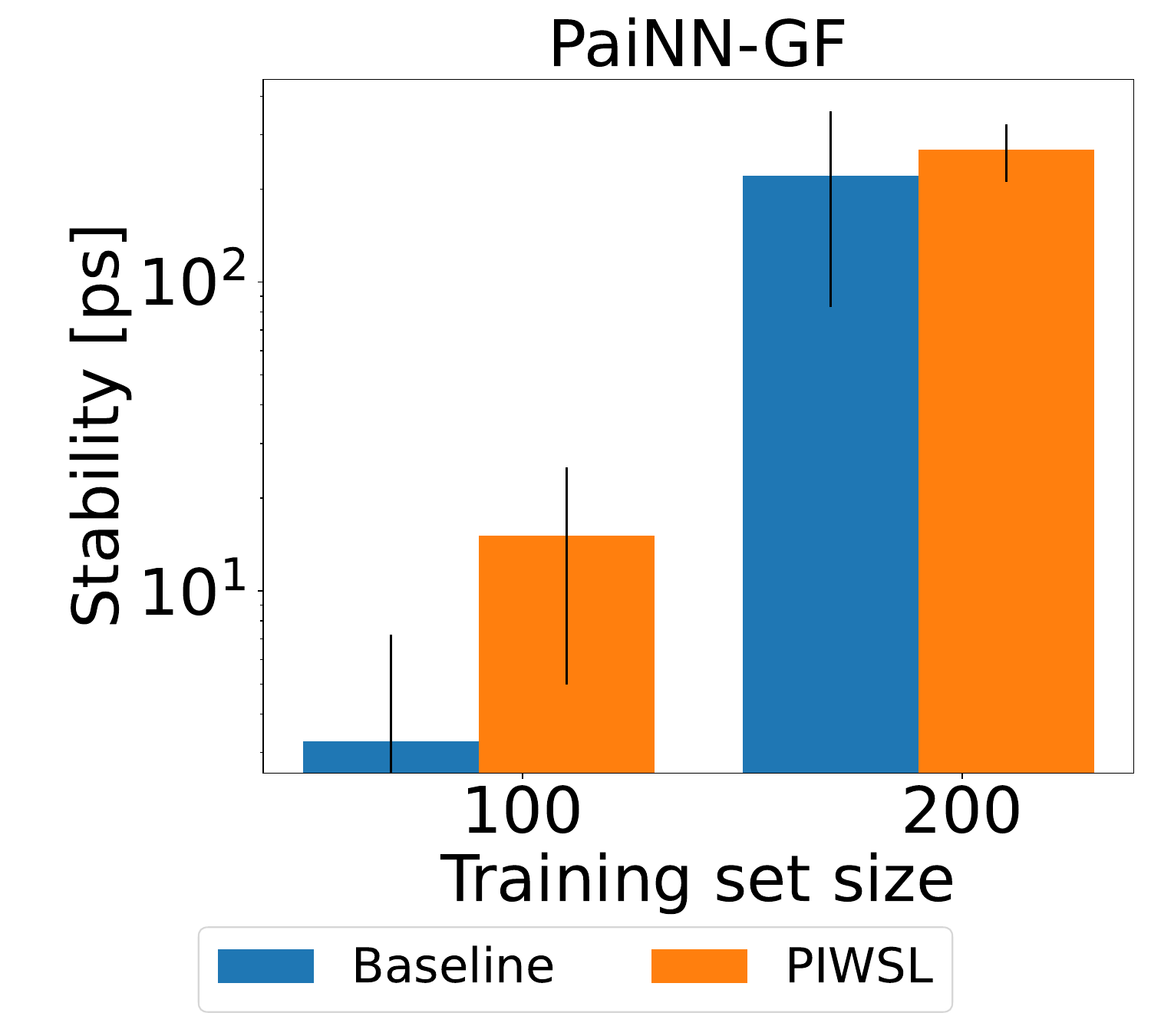}    
    \caption{\textbf{Stability analysis of the MLIP models during MD simulations.} Stability during MD simulations is assessed for the baseline MLIP models and those trained with PIWSL. Left: Models with the direct force branch. Right: Models with forces computed as negative gradients of the energy. All results are obtained for the aspirin molecule and MD simulations in the microcanonical (NVE) statistical ensemble. We measure stability during MD simulations according to \citet{fu2023forces}.}
    \label{fig:stability}
\end{figure*}

We evaluate the prediction variance and robustness of an MLIP model trained with PIWSL using the aspirin molecule, focusing on the potential energy's dependence on the C--H bond length. In this work, robustness refers to the prediction robustness of an MLIP to perturbations in atomic coordinates. In the literature, the robustness of MLIPs also means their stability during MD simulations. We train PaiNN on the rMD17 aspirin data set using $100$ and $200$ configurations with and without the PIWSL loss. The detailed training setup and errors of the used MLIPs are summarized in \secref{sec:vis-setup}. We examine the potential energy varying the length of a C--H bond from $0.9$ \AA{} to $1.4$ \AA. The equilibrium C--H bond length is about $1.09$ \AA. The results in \figref{fig:enter-label} (c) and (d) demonstrate that the PIWSL method improves the predicted potential energy profile, indicating improved robustness under perturbations of atom coordinates.

Although the estimated potential energies do not always match the reference values, the direction between the original and perturbed configurations, indicated by arrows in \figref{fig:enter-label}, consistently follows the gradient of the reference potential energy, corresponding to the negative force. In \figref{fig:enter-label}, we use a perturbation length of $||\dr|| = 0.01$~\AA. This consistency with the energy gradient underscores the PIWSL method's effectiveness, ensuring the alignment between predicted total energies and force and improving the corresponding RMSE values. As discussed in \secref{sec:conservative-force}, PIWSL also addresses the limitation of MLIPs that employ separate force branches and do not guarantee the prediction of conservative forces. The proposed method reduces the curl of predicted forces, as detailed in \secref{sec:force-rot-experiment}, although complete elimination of the curl remains a challenge. In summary, PIWSL minimizes individual energy and force errors, improving the overall accuracy of MLIPs.

To further assess PIWSL's impact, we evaluate the robustness during MD simulations of the MLIP models trained with and without PIWSL. We consider MD simulations of the aspirin molecule, with corresponding results presented in \figref{fig:stability}. We measured stability following the approach proposed in \citet{fu2023forces}. A detailed experimental setup is provided in \secref{sec:md-setup}. The results demonstrate that PIWSL improves the stability of MD simulations for both the direct and gradient-based force prediction models. The simulation times in \figref{fig:stability} are shorter than those reported by \citet{fu2023forces}. This difference arises from our choice to perform MD simulations in the microcanonical (NVE) statistical ensemble instead of the canonical (NVT) statistical ensemble in \citet{fu2023forces} to assess stability more accurately without the influence of a thermostat. Results for MD simulations conducted in the canonical statistical ensemble are provided in \secref{sec:md-setup}. 

\subsection{Fine-Tuning of Foundation Models\label{sec:md17cc-finetune}}

In this section, we present a case study demonstrating the application of PIWSL to the fine-tuning of foundation models using a sparse data set. The detailed experimental setup is provided in \secref{sec:md17cc-setup}. First, we consider fine-tuning MACE-OFF (large) \citep{Kovacs2023}, a MACE model pre-trained on the SPICE data set\citep{eastman2023spice}, including QMugs and liquid water subsets \citep{isert2022qmugs, schran2021machine}. The foundation model is fine-tuned using the aspirin data set with energies and forces evaluated at the CCSD level of theory \citep{chmiela2018towards}. We consider a more challenging scenario, e.g., where CC energies are extrapolated to the CBS limit. Thus, we use only energy labels for training.

We follow the fine-tuning procedure outlined in the official repository\footnote{\url{https://mace-docs.readthedocs.io/en/latest/guide/finetuning.html}}, adjusting the learning rate from $10^{-2}$ to $10^{-3}$ to enhance final accuracy of the model. The results, presented in \tabref{tb:mace_md17cc_results}, show that PIWSL significantly improves model accuracy, achieving approximately 40~\% lower RMSE values. Notably, the model trained with PIWSL on 512 samples performs similarly to the model trained without it using nearly doubled training set size, i.e., 950 configurations. These results highlight the data efficiency of our approach. Furthermore, both fine-tuned MACE-OFF models outperform models trained from scratch.

\begin{table}[t!]
    \captionof{table}{\textbf{Results for models trained on the MD17(CCSD) data set without reference atomic forces.} All models are trained on aspirin data without force labels. Energy RMSE is given in kcal/mol, while force RMSE is in kcal/mol/\AA. \label{tb:mace_md17cc_results}}
    \begin{center}
    \resizebox{\columnwidth}{!}{
    \begin{tabular}{lcccrr}
    \toprule
    Model                        & Samples & Epoch &   & Baseline          & PIWSL                     \\
    \cmidrule(lr){1-4} \cmidrule(lr){5-6}
    \multirow[c]{2}{*}{MACE-OFF} & 950 & 100  & E & 1.21 $\pm$ 0.00 & {\bf 0.72 $\pm$ 0.01}    \\
                                   &    &  & F\textsuperscript{\emph{a}} & 6.90 $\pm$ 0.01 & {\bf 3.77 $\pm$ 0.13}    \\
    \cmidrule(lr){1-4} \cmidrule(lr){5-6}                                                            
    \multirow[c]{2}{*}{MACE-OFF} & 512& 100  & E & 2.03 $\pm$ 0.03 & {\bf 1.21 $\pm$ 0.02} \\
                          &      & & F\textsuperscript{\emph{a}} & 10.96 $\pm$ 0.13 & {\bf 6.55 $\pm$ 0.27}   \\
    \cmidrule(lr){1-4} \cmidrule(lr){5-6}                                                            
    \multirow[c]{2}{*}{MACE (scratch)} & 950 & 1000 & E & 3.27 $\pm$ 0.16 & {\bf 2.10 $\pm$ 0.30} \\
                                       &     & & F\textsuperscript{\emph{a}} & 18.05 $\pm$ 0.82 & {\bf 10.99 $\pm$ 0.18}   \\
    \bottomrule
    \end{tabular}
    }
    \end{center}
    \footnotesize{\textsuperscript{\emph{a}} We evaluated the accuracy of predicted atomic forces using corresponding force labels provided in the test data set.}
\end{table}

We also evaluate the impact of PIWSL on data sets containing conformations of a single large molecule. For this purpose, we have chosen the buckyball catcher molecule from the MD22 data set \citep{md22} with 148 atoms. We used two foundation models: MACE-MP (large) \citep{Batatia2023} and MACE-OFF (large) \citep{Kovacs2023}. We considered fine-tuning on a sparse data set; thus, both models were trained using only 50 samples. The results, presented in \Tabref{tb:MD22_buckyball_results_FM}, demonstrate again that PIWSL improves the fine-tuning of foundation models. However, the performance of MACE-MP is worse compared to MACE-OFF, highlighting the importance of careful selection of the pre-trained model~\citep{Kjeldal_2024}. Note that a more rigorous hyper-parameter search could further enhance the performance of PIWSL. However, this is beyond the scope of this paper, as our primary objective is to demonstrate the relative effectiveness of the PIWSL approach.

\begin{table}
    \begin{center}
    \caption{\textbf{Energy (E) and force (F) mean-absolute errors (MAEs) for MACE, MACE-OFF, and MACE-MP models fine-tuned on the MD22 data set.} All models are fine-tuned using 50 training samples of the buckyball-catcher molecule from MD22. Energy MAE is given in kcal/mol, while force MAE is in kcal/mol/\AA. \label{tb:MD22_buckyball_results_FM}}
    \resizebox{\columnwidth}{!}{
    \begin{tabular}{lrlrr}
    \toprule
    Model            & Train Epoch                 &   & Baseline              & PIWSL                  \\
    \cmidrule(lr){1-3} \cmidrule(lr){4-5}
    \multirow[c]{2}{*}{MACE (scratch)} & 800 & E & 1.046 $\pm$ 0.095  & {\bf 0.745 $\pm$ 0.031}  \\
                                       &     & F & 0.294 $\pm$ 0.002 & {\bf 0.290 $\pm$ 0.002} \\
    \cmidrule(lr){1-3} \cmidrule(lr){4-5}
    \multirow[c]{2}{*}{MACE-MP}        & 100 & E & {\bf 2.111 $\pm$ 0.453} & {\bf 2.033 $\pm$ 0.149}  \\
                                       &     & F & 0.716 $\pm$ 0.009& {\bf 0.700 $\pm$ 0.011} \\
    \cmidrule(lr){1-3} \cmidrule(lr){4-5} 
     \multirow[c]{2}{*}{MACE-OFF}    & 100 & E & 1.159 $\pm$ 0.154 & {\bf 0.992 $\pm$ 0.051}  \\
                                       &     & F & 0.346 $\pm$ 0.003 & {\bf 0.337 $\pm$ 0.001} \\
    \bottomrule
    \end{tabular}}
    \end{center}
\end{table}

\section{Discussion and Limitations \label{sec:conlcusion}}

This work introduces the PIWSL method, encompassing two distinct physics-informed weakly supervised loss functions, for learning MLIPs. These losses provide the physics-informed weak labels based on the Taylor expansion (PITC loss) and the spatial consistency (PISC loss) of the potential energy. These physics-informed weak labels enable any MLIP to improve its accuracy and robustness, particularly in scenarios characterized by sparse training data, which are common when investigating a new molecular or material system. The improved accuracy and robustness of MLIPs can allow running sufficiently long MD simulations, resulting in a more effective use of active learning approaches. Our extensive experiments demonstrate notable efficacy and efficiency of our method from various aspects: (i) dependence on the training data set size, (ii) the potential energy prediction variance and robustness in terms of a perturbation on a C--H bond length as well as robustness during MD simulations, and (iii) application to the fine-tuning of foundation models. In particular, it is shown that our PIWSL method enables MLIPs to improve the accuracy of their force predictions even without force labels used explicitly during training. Therefore, PIWSL opens a new possibility for training MLIPs using highly accurate reference methods, such as CCSD(T)/CBS.

\textbf{Limitations.} The proposed PIWSL method is tailored to ML models that predict atomic forces and total energies of atomic systems. It cannot be applied to other ML problems unrelated to computational chemistry or materials science. Although this work uses up to the second-order Taylor expansion to obtain weak labels in \eqref{eq:taylor-wsl}, employing more sophisticated higher-order ordinary differential equations is a viable alternative.


\newpage

\section*{Acknowledgements}
MN acknowledges support from the Deutsche Forschungsgemeinschaft (DFG, German Research Foundation) under Germany's Excellence Strategy - EXC 2075 – 390740016 and the Stuttgart Center for Simulation Science (SimTech).

\bibliography{reference}

\begin{thebibliography}{69}
\providecommand{\natexlab}[1]{#1}
\providecommand{\url}[1]{\texttt{#1}}
\expandafter\ifx\csname urlstyle\endcsname\relax
  \providecommand{\doi}[1]{doi: #1}\else
  \providecommand{\doi}{doi: \begingroup \urlstyle{rm}\Url}\fi

\bibitem[Akiba et~al.(2019)Akiba, Sano, Yanase, Ohta, and Koyama]{optuna_2019}
Akiba, T., Sano, S., Yanase, T., Ohta, T., and Koyama, M.
\newblock Optuna: A next-generation hyperparameter optimization framework.
\newblock \emph{Int. Conf. Knowl. Discov. Data Min.}, 2019.

\bibitem[Artrith \& Urban(2016)Artrith and Urban]{ARTRITH2016135}
Artrith, N. and Urban, A.
\newblock An implementation of artificial neural-network potentials for atomistic materials simulations: Performance for tio2.
\newblock \emph{Comput. Mater. Sci.}, 114:\penalty0 135--150, 2016.
\newblock ISSN 0927--0256.
\newblock \doi{https://doi.org/10.1016/j.commatsci.2015.11.047}.

\bibitem[Artrith et~al.(2011)Artrith, Morawietz, and Behler]{artrith2011high}
Artrith, N., Morawietz, T., and Behler, J.
\newblock High-dimensional neural-network potentials for multicomponent systems: Applications to zinc oxide.
\newblock \emph{Phys. Rev. B}, 83:\penalty0 153101, 2011.

\bibitem[Bartlett \& Musia\l{}(2007)Bartlett and Musia\l{}]{Bartlett2007}
Bartlett, R.~J. and Musia\l{}, M.
\newblock Coupled-cluster theory in quantum chemistry.
\newblock \emph{Rev. Mod. Phys.}, 79:\penalty0 291--352, 2007.
\newblock \doi{10.1103/RevModPhys.79.291}.

\bibitem[Batatia et~al.(2022)Batatia, Kovacs, Simm, Ortner, and Csanyi]{batatia2022mace}
Batatia, I., Kovacs, D.~P., Simm, G. N.~C., Ortner, C., and Csanyi, G.
\newblock {{MACE}: Higher Order Equivariant Message Passing Neural Networks for Fast and Accurate Force Fields}.
\newblock \emph{Adv. Neural Inf. Process. Syst.}, 35:\penalty0 11423--11436, 2022.
\newblock URL \url{https://openreview.net/forum?id=YPpSngE-ZU}.

\bibitem[Batatia et~al.(2023)Batatia, Benner, Chiang, Elena, Kovács, Riebesell, Advincula, Asta, Baldwin, Bernstein, Bhowmik, Blau, Cărare, Darby, De, Pia, Deringer, Elijošius, El-Machachi, Fako, Ferrari, Genreith-Schriever, George, Goodall, Grey, Han, Handley, Heenen, Hermansson, Holm, Jaafar, Hofmann, Jakob, Jung, Kapil, Kaplan, Karimitari, Kroupa, Kullgren, Kuner, Kuryla, Liepuoniute, Margraf, Magdău, Michaelides, Moore, Naik, Niblett, Norwood, O'Neill, Ortner, Persson, Reuter, Rosen, Schaaf, Schran, Sivonxay, Stenczel, Svahn, Sutton, van~der Oord, Varga-Umbrich, Vegge, Vondrák, Wang, Witt, Zills, and Csányi]{Batatia2023}
Batatia, I., Benner, P., Chiang, Y., Elena, A.~M., Kovács, D.~P., Riebesell, J., Advincula, X.~R., Asta, M., Baldwin, W.~J., Bernstein, N., Bhowmik, A., Blau, S.~M., Cărare, V., Darby, J.~P., De, S., Pia, F.~D., Deringer, V.~L., Elijošius, R., El-Machachi, Z., Fako, E., Ferrari, A.~C., Genreith-Schriever, A., George, J., Goodall, R. E.~A., Grey, C.~P., Han, S., Handley, W., Heenen, H.~H., Hermansson, K., Holm, C., Jaafar, J., Hofmann, S., Jakob, K.~S., Jung, H., Kapil, V., Kaplan, A.~D., Karimitari, N., Kroupa, N., Kullgren, J., Kuner, M.~C., Kuryla, D., Liepuoniute, G., Margraf, J.~T., Magdău, I.-B., Michaelides, A., Moore, J.~H., Naik, A.~A., Niblett, S.~P., Norwood, S.~W., O'Neill, N., Ortner, C., Persson, K.~A., Reuter, K., Rosen, A.~S., Schaaf, L.~L., Schran, C., Sivonxay, E., Stenczel, T.~K., Svahn, V., Sutton, C., van~der Oord, C., Varga-Umbrich, E., Vegge, T., Vondrák, M., Wang, Y., Witt, W.~C., Zills, F., and Csányi, G.
\newblock A foundation model for atomistic materials chemistry, 2023.

\bibitem[Batzner et~al.(2022)Batzner, Musaelian, Sun, Geiger, Mailoa, Kornbluth, Molinari, Smidt, and Kozinsky]{batzner2023}
Batzner, S., Musaelian, A., Sun, L., Geiger, M., Mailoa, J.~P., Kornbluth, M., Molinari, N., Smidt, T.~E., and Kozinsky, B.
\newblock E(3)-equivariant graph neural networks for data-efficient and accurate interatomic potentials.
\newblock \emph{Nat. Commun.}, 13:\penalty0 2453, 2022.

\bibitem[Behler \& Parrinello(2007)Behler and Parrinello]{behler2007generalized}
Behler, J. and Parrinello, M.
\newblock Generalized neural-network representation of high-dimensional potential-energy surfaces.
\newblock \emph{Phys. Rev. Lett.}, 98:\penalty0 146401, 2007.

\bibitem[Blank et~al.(1995)Blank, Brown, Calhoun, and Doren]{blank1995neural}
Blank, T.~B., Brown, S.~D., Calhoun, A.~W., and Doren, D.~J.
\newblock Neural network models of potential energy surfaces.
\newblock \emph{J. Chem. Phys.}, 103:\penalty0 4129--4137, 1995.

\bibitem[Briganti \& Lunghi(2023)Briganti and Lunghi]{Briganti2023}
Briganti, V. and Lunghi, A.
\newblock Efficient generation of stable linear machine-learning force fields with uncertainty-aware active learning.
\newblock \emph{Mach. Learn.: Sci. Technol.}, 4:\penalty0 035005, jul 2023.
\newblock \doi{10.1088/2632-2153/ace418}.
\newblock URL \url{https://dx.doi.org/10.1088/2632-2153/ace418}.

\bibitem[Cai et~al.(2022)Cai, Mao, Wang, Yin, and Karniadakis]{cai2022physics}
Cai, S., Mao, Z., Wang, Z., Yin, M., and Karniadakis, G.~E.
\newblock Physics-informed neural networks (pinns) for fluid mechanics: A review.
\newblock \emph{Acta Mech. Sin.}, pp.\  1--12, 2022.

\bibitem[Carloni et~al.(2002)Carloni, Rothlisberger, and Parrinello]{Carloni2002}
Carloni, P., Rothlisberger, U., and Parrinello, M.
\newblock The role and perspective of ab initio molecular dynamics in the study of biological systems.
\newblock \emph{Acc. Chem. Res.}, 35:\penalty0 455--464, 2002.
\newblock \doi{10.1021/ar010018u}.

\bibitem[Chanussot et~al.(2021)Chanussot, Das, Goyal, Lavril, Shuaibi, Riviere, Tran, Heras-Domingo, Ho, Hu, Palizhati, Sriram, Wood, Yoon, Parikh, Zitnick, and Ulissi]{ocp_dataset}
Chanussot, L., Das, A., Goyal, S., Lavril, T., Shuaibi, M., Riviere, M., Tran, K., Heras-Domingo, J., Ho, C., Hu, W., Palizhati, A., Sriram, A., Wood, B., Yoon, J., Parikh, D., Zitnick, C.~L., and Ulissi, Z.
\newblock {Open Catalyst 2020 (OC20) Dataset and Community Challenges}.
\newblock \emph{ACS Catal.}, 11:\penalty0 6059--6072, 2021.
\newblock \doi{10.1021/acscatal.0c04525}.

\bibitem[Chmiela et~al.(2017)Chmiela, Tkatchenko, Sauceda, Poltavsky, Sch{\"u}tt, and M{\"u}ller]{chmiela2017machine}
Chmiela, S., Tkatchenko, A., Sauceda, H.~E., Poltavsky, I., Sch{\"u}tt, K.~T., and M{\"u}ller, K.-R.
\newblock Machine learning of accurate energy-conserving molecular force fields.
\newblock \emph{Sci. Adv.}, 3:\penalty0 e1603015, 2017.

\bibitem[Chmiela et~al.(2018)Chmiela, Sauceda, M{\"u}ller, and Tkatchenko]{chmiela2018towards}
Chmiela, S., Sauceda, H.~E., M{\"u}ller, K.-R., and Tkatchenko, A.
\newblock Towards exact molecular dynamics simulations with machine-learned force fields.
\newblock \emph{Nat. Commun.}, 9:\penalty0 3887, 2018.

\bibitem[Chmiela et~al.(2023)Chmiela, Vassilev-Galindo, Unke, Kabylda, Sauceda, Tkatchenko, and M{\"u}ller]{md22}
Chmiela, S., Vassilev-Galindo, V., Unke, O.~T., Kabylda, A., Sauceda, H.~E., Tkatchenko, A., and M{\"u}ller, K.-R.
\newblock Accurate global machine learning force fields for molecules with hundreds of atoms.
\newblock \emph{Science Advances}, 9\penalty0 (2):\penalty0 eadf0873, 2023.

\bibitem[Christensen \& von Lilienfeld(2020)Christensen and von Lilienfeld]{Christensen2020}
Christensen, A.~S. and von Lilienfeld, O.~A.
\newblock On the role of gradients for machine learning of molecular energies and forces.
\newblock \emph{Mach. Learn.: Sci. Technol.}, 1:\penalty0 045018, oct 2020.
\newblock \doi{10.1088/2632-2153/abba6f}.

\bibitem[Cooper et~al.(2020)Cooper, K{\"a}stner, Urban, and Artrith]{cooper2020efficient}
Cooper, A.~M., K{\"a}stner, J., Urban, A., and Artrith, N.
\newblock Efficient training of ann potentials by including atomic forces via taylor expansion and application to water and a transition-metal oxide.
\newblock \emph{npj Comput. Mater.}, 6:\penalty0 54, 2020.

\bibitem[Drautz(2019)]{drautz2019atomic}
Drautz, R.
\newblock Atomic cluster expansion for accurate and transferable interatomic potentials.
\newblock \emph{Phys. Rev. B}, 99:\penalty0 014104, 2019.

\bibitem[Eastman et~al.(2023)Eastman, Behara, Dotson, Galvelis, Herr, Horton, Mao, Chodera, Pritchard, Wang, et~al.]{eastman2023spice}
Eastman, P., Behara, P.~K., Dotson, D.~L., Galvelis, R., Herr, J.~E., Horton, J.~T., Mao, Y., Chodera, J.~D., Pritchard, B.~P., Wang, Y., et~al.
\newblock Spice, a dataset of drug-like molecules and peptides for training machine learning potentials.
\newblock \emph{Scientific Data}, 10\penalty0 (1):\penalty0 11, 2023.

\bibitem[Feller et~al.(2006)Feller, Peterson, and Crawford]{Feller2006}
Feller, D., Peterson, K.~A., and Crawford, T.~D.
\newblock Sources of error in electronic structure calculations on small chemical systems.
\newblock \emph{J. Chem. Phys.}, 124:\penalty0 054107, 2006.
\newblock \doi{10.1063/1.2137323}.

\bibitem[Fu et~al.(2023)Fu, Wu, Wang, Xie, Keten, Gomez-Bombarelli, and Jaakkola]{fu2023forces}
Fu, X., Wu, Z., Wang, W., Xie, T., Keten, S., Gomez-Bombarelli, R., and Jaakkola, T.~S.
\newblock Forces are not enough: Benchmark and critical evaluation for machine learning force fields with molecular simulations.
\newblock \emph{Transact. Mach. Learn. Res.}, 2023.

\bibitem[Godwin et~al.(2022)Godwin, Schaarschmidt, Gaunt, Sanchez-Gonzalez, Rubanova, Veli{\v{c}}kovi{\'c}, Kirkpatrick, and Battaglia]{godwin2021simple}
Godwin, J., Schaarschmidt, M., Gaunt, A.~L., Sanchez-Gonzalez, A., Rubanova, Y., Veli{\v{c}}kovi{\'c}, P., Kirkpatrick, J., and Battaglia, P.
\newblock Simple {GNN} regularisation for 3d molecular property prediction and beyond.
\newblock \emph{Int. Conf. Learn. Represent.}, 2022.

\bibitem[Goodfellow et~al.(2014)Goodfellow, Shlens, and Szegedy]{goodfellow2014explaining}
Goodfellow, I.~J., Shlens, J., and Szegedy, C.
\newblock Explaining and harnessing adversarial examples.
\newblock \emph{Int. Conf. Learn. Represent.}, 2014.

\bibitem[Hobza \& Šponer(2002)Hobza and Šponer]{Hobza2002}
Hobza, P. and Šponer, J.
\newblock Toward true dna base-stacking energies: Mp2, ccsd(t), and complete basis set calculations.
\newblock \emph{J. Am. Chem. Soc.}, 124:\penalty0 11802--11808, 2002.
\newblock \doi{10.1021/ja026759n}.

\bibitem[Hohenberg \& Kohn(1964)Hohenberg and Kohn]{Hohenberg1964}
Hohenberg, P. and Kohn, W.
\newblock Inhomogeneous electron gas.
\newblock \emph{Phys. Rev.}, 136:\penalty0 B864--B871, 1964.
\newblock \doi{10.1103/PhysRev.136.B864}.

\bibitem[Hoover(1985)]{hoover1985canonical}
Hoover, W.~G.
\newblock Canonical dynamics: Equilibrium phase-space distributions.
\newblock \emph{Physical review A}, 31\penalty0 (3):\penalty0 1695, 1985.

\bibitem[Hu et~al.(2021)Hu, Shuaibi, Das, Goyal, Sriram, Leskovec, Parikh, and Zitnick]{hu2021forcenet}
Hu, W., Shuaibi, M., Das, A., Goyal, S., Sriram, A., Leskovec, J., Parikh, D., and Zitnick, C.~L.
\newblock Forcenet: A graph neural network for large-scale quantum calculations.
\newblock \emph{Int. Conf. Learn. Represent}, 2021.

\bibitem[Iftimie et~al.(2005)Iftimie, Minary, and Tuckerman]{Iftimie2005}
Iftimie, R., Minary, P., and Tuckerman, M.~E.
\newblock Ab initio molecular dynamics: Concepts, recent developments, and future trends.
\newblock \emph{Proc. Natl. Acad. Sci.}, 102:\penalty0 6654--6659, 2005.
\newblock \doi{10.1073/pnas.0500193102}.

\bibitem[Isert et~al.(2022)Isert, Atz, Jim{\'e}nez-Luna, and Schneider]{isert2022qmugs}
Isert, C., Atz, K., Jim{\'e}nez-Luna, J., and Schneider, G.
\newblock Qmugs, quantum mechanical properties of drug-like molecules.
\newblock \emph{Scientific Data}, 9\penalty0 (1):\penalty0 273, 2022.

\bibitem[Kjeldal \& Eriksen(2024)Kjeldal and Eriksen]{Kjeldal_2024}
Kjeldal, F. and Eriksen, J.~J.
\newblock Transferability of atom-based neural networks.
\newblock \emph{Machine Learning: Science and Technology}, 5\penalty0 (4):\penalty0 045059, dec 2024.
\newblock \doi{10.1088/2632-2153/ad9709}.
\newblock URL \url{https://dx.doi.org/10.1088/2632-2153/ad9709}.

\bibitem[Kohn \& Sham(1965)Kohn and Sham]{Kohn1965}
Kohn, W. and Sham, L.~J.
\newblock {Self-Consistent Equations Including Exchange and Correlation Effects}.
\newblock \emph{Phys. Rev.}, 140:\penalty0 A1133--A1138, 1965.
\newblock \doi{10.1103/PhysRev.140.A1133}.

\bibitem[Kovács et~al.(2023)Kovács, Moore, Browning, Batatia, Horton, Kapil, Witt, Magdău, Cole, and Csányi]{Kovacs2023}
Kovács, D.~P., Moore, J.~H., Browning, N.~J., Batatia, I., Horton, J.~T., Kapil, V., Witt, W.~C., Magdău, I.-B., Cole, D.~J., and Csányi, G.
\newblock Mace-off23: Transferable machine learning force fields for organic molecules, 2023.

\bibitem[Li et~al.(2015)Li, Kermode, and De~Vita]{Li2015}
Li, Z., Kermode, J.~R., and De~Vita, A.
\newblock Molecular dynamics with on-the-fly machine learning of quantum-mechanical forces.
\newblock \emph{Phys. Rev. Lett.}, 114:\penalty0 096405, 2015.

\bibitem[Liao et~al.(2023)Liao, Wood, Das, and Smidt]{equiformer_v2}
Liao, Y.-L., Wood, B., Das, A., and Smidt, T.
\newblock {EquiformerV2: Improved Equivariant Transformer for Scaling to Higher-Degree Representations}.
\newblock \emph{Int. Conf. Learn. Represent.}, 2023.

\bibitem[Liao et~al.(2024)Liao, Smidt, Shuaibi*, and Das*]{DeNS}
Liao, Y.-L., Smidt, T., Shuaibi*, M., and Das*, A.
\newblock Generalizing denoising to non-equilibrium structures improves equivariant force fields.
\newblock \emph{arXiv preprint arXiv:2403.09549}, 2024.

\bibitem[Merchant et~al.(2023)Merchant, Batzner, Schoenholz, Aykol, Cheon, and Cubuk]{Merchant2023}
Merchant, A., Batzner, S., Schoenholz, S.~S., Aykol, M., Cheon, G., and Cubuk, E.~D.
\newblock Scaling deep learning for materials discovery.
\newblock \emph{Nature}, 624:\penalty0 80--85, 2023.
\newblock \doi{10.1038/s41586-023-06735-9}.

\bibitem[Miyato et~al.(2018)Miyato, Maeda, Koyama, and Ishii]{miyato2018virtual}
Miyato, T., Maeda, S.-i., Koyama, M., and Ishii, S.
\newblock Virtual adversarial training: a regularization method for supervised and semi-supervised learning.
\newblock \emph{IEEE Trans. Pattern Anal. Mach. Intell.}, 41:\penalty0 1979--1993, 2018.

\bibitem[Musaelian et~al.(2023)Musaelian, Batzner, Johansson, Sun, Owen, Kornbluth, and Kozinsky]{musaelian2023learning}
Musaelian, A., Batzner, S., Johansson, A., Sun, L., Owen, C.~J., Kornbluth, M., and Kozinsky, B.
\newblock Learning local equivariant representations for large-scale atomistic dynamics.
\newblock \emph{Nat. Commun.}, 14:\penalty0 579, 2023.

\bibitem[Ni et~al.(2024)Ni, Feng, Ma, Ma, and Lan]{ni2023sliced}
Ni, Y., Feng, S., Ma, W.-Y., Ma, Z.-M., and Lan, Y.
\newblock Sliced denoising: A physics-informed molecular pre-training method.
\newblock \emph{Int. Conf. Learn. Represent.}, 2024.

\bibitem[Nos{\'e}(1984)]{nose1984unified}
Nos{\'e}, S.
\newblock A unified formulation of the constant temperature molecular dynamics methods.
\newblock \emph{The Journal of chemical physics}, 81\penalty0 (1):\penalty0 511--519, 1984.

\bibitem[Parrinello(1997)]{Parrinello1997}
Parrinello, M.
\newblock From silicon to rna: The coming of age of ab initio molecular dynamics.
\newblock \emph{Solid State Commun.}, 102:\penalty0 107--120, 1997.
\newblock ISSN 0038--1098.
\newblock \doi{https://doi.org/10.1016/S0038-1098(96)00723-5}.

\bibitem[Passaro \& Zitnick(2023)Passaro and Zitnick]{escn}
Passaro, S. and Zitnick, C.~L.
\newblock {Reducing SO(3) Convolutions to SO(2) for Efficient Equivariant GNNs}.
\newblock \emph{Int. Conf. Mach. Learn.}, 202:\penalty0 27420--27438, 2023.

\bibitem[Podryabinkin \& Shapeev(2017)Podryabinkin and Shapeev]{Podryabinkin2017}
Podryabinkin, E.~V. and Shapeev, A.~V.
\newblock Active learning of linearly parametrized interatomic potentials.
\newblock \emph{Comput. Mater. Sci.}, 140:\penalty0 171--180, 2017.
\newblock \doi{https://doi.org/10.1016/j.commatsci.2017.08.031}.

\bibitem[Purvis \& Bartlett(1982)Purvis and Bartlett]{Purvis1982}
Purvis, G.~D. and Bartlett, R.~J.
\newblock A full coupled‐cluster singles and doubles model: The inclusion of disconnected triples.
\newblock \emph{J. Chem. Phys.}, 76:\penalty0 1910--1918, 1982.
\newblock \doi{10.1063/1.443164}.

\bibitem[{Raissi} et~al.(2019){Raissi}, {Perdikaris}, and {Karniadakis}]{2019JCoPh.378..686R}
{Raissi}, M., {Perdikaris}, P., and {Karniadakis}, G.~E.
\newblock {Physics-informed neural networks: A deep learning framework for solving forward and inverse problems involving nonlinear partial differential equations}.
\newblock \emph{J. Comput. Phys.}, 378:\penalty0 686--707, 2019.
\newblock \doi{10.1016/j.jcp.2018.10.045}.

\bibitem[Schran et~al.(2021)Schran, Thiemann, Rowe, M{\"u}ller, Marsalek, and Michaelides]{schran2021machine}
Schran, C., Thiemann, F.~L., Rowe, P., M{\"u}ller, E.~A., Marsalek, O., and Michaelides, A.
\newblock Machine learning potentials for complex aqueous systems made simple.
\newblock \emph{Proceedings of the National Academy of Sciences}, 118\penalty0 (38):\penalty0 e2110077118, 2021.

\bibitem[Sch\"{u}tt et~al.(2017)Sch\"{u}tt, Kindermans, Sauceda~Felix, Chmiela, Tkatchenko, and M\"{u}ller]{schutt2017schnet}
Sch\"{u}tt, K., Kindermans, P.-J., Sauceda~Felix, H.~E., Chmiela, S., Tkatchenko, A., and M\"{u}ller, K.-R.
\newblock Schnet: A continuous-filter convolutional neural network for modeling quantum interactions.
\newblock \emph{Adv. Neural Inf. Process. Syst.}, 30:\penalty0 991--1001, 2017.

\bibitem[Sch{\"u}tt et~al.(2021)Sch{\"u}tt, Unke, and Gastegger]{schutt2021equivariant}
Sch{\"u}tt, K.~T., Unke, O.~T., and Gastegger, M.
\newblock Equivariant message passing for the prediction of tensorial properties and molecular spectra.
\newblock \emph{Int. Conf. Mach. Learn.}, 139:\penalty0 9377--9388, 2021.

\bibitem[Shapeev(2016)]{Shapeev2016}
Shapeev, A.~V.
\newblock Moment tensor potentials: A class of systematically improvable interatomic potentials.
\newblock \emph{Multiscale Model. Simul.}, 14:\penalty0 1153--1173, 2016.
\newblock \doi{10.1137/15M1054183}.

\bibitem[Shuaibi et~al.(2021{\natexlab{a}})Shuaibi, Kolluru, Das, Grover, Sriram, Ulissi, and Zitnick]{shuaibi2021rotation}
Shuaibi, M., Kolluru, A., Das, A., Grover, A., Sriram, A., Ulissi, Z., and Zitnick, C.~L.
\newblock Rotation invariant graph neural networks using spin convolutions, 2021{\natexlab{a}}.

\bibitem[Shuaibi et~al.(2021{\natexlab{b}})Shuaibi, Sivakumar, Chen, and Ulissi]{Shuaibi2021}
Shuaibi, M., Sivakumar, S., Chen, R.~Q., and Ulissi, Z.~W.
\newblock Enabling robust offline active learning for machine learning potentials using simple physics-based priors.
\newblock \emph{Mach. Learn.: Sci. Technol.}, 2:\penalty0 025007, dec 2021{\natexlab{b}}.
\newblock \doi{10.1088/2632-2153/abcc44}.
\newblock URL \url{https://dx.doi.org/10.1088/2632-2153/abcc44}.

\bibitem[Smith et~al.(2017)Smith, Isayev, and Roitberg]{Smith2017}
Smith, J.~S., Isayev, O., and Roitberg, A.~E.
\newblock {ANI-1: An extensible neural network potential with DFT accuracy at force field computational cost}.
\newblock \emph{Chem. Sci.}, 8:\penalty0 3192--3203, 2017.
\newblock \doi{10.1039/C6SC05720A}.

\bibitem[Smith et~al.(2019)Smith, Nebgen, Zubatyuk, Lubbers, Devereux, Barros, Tretiak, Isayev, and Roitberg]{Smith2019}
Smith, J.~S., Nebgen, B.~T., Zubatyuk, R., Lubbers, N., Devereux, C., Barros, K., Tretiak, S., Isayev, O., and Roitberg, A.~E.
\newblock {Approaching coupled cluster accuracy with a general-purpose neural network potential through transfer learning}.
\newblock \emph{Nat. Commun.}, 10:\penalty0 2903, 2019.
\newblock ISSN 2041--1723.
\newblock \doi{10.1038/s41467-019-10827-4}.

\bibitem[Smith et~al.(2020)Smith, Zubatyuk, Nebgen, Lubbers, Barros, Roitberg, Isayev, and Tretiak]{smith2020ani}
Smith, J.~S., Zubatyuk, R., Nebgen, B., Lubbers, N., Barros, K., Roitberg, A.~E., Isayev, O., and Tretiak, S.
\newblock The ani-1ccx and ani-1x data sets, coupled-cluster and density functional theory properties for molecules.
\newblock \emph{Sci. Data}, 7:\penalty0 134, 2020.

\bibitem[Thomas et~al.(2018)Thomas, Smidt, Kearnes, Yang, Li, Kohlhoff, and Riley]{thomas2018tensor}
Thomas, N., Smidt, T., Kearnes, S., Yang, L., Li, L., Kohlhoff, K., and Riley, P.
\newblock Tensor field networks: Rotation- and translation-equivariant neural networks for 3d point clouds, 2018.

\bibitem[Unke \& Meuwly(2019)Unke and Meuwly]{Unke2019}
Unke, O.~T. and Meuwly, M.
\newblock Physnet: A neural network for predicting energies, forces, dipole moments, and partial charges.
\newblock \emph{J. Chem. Theory Comput.}, 15:\penalty0 3678--3693, 2019.

\bibitem[Unke et~al.(2021)Unke, Chmiela, Sauceda, Gastegger, Poltavsky, Sch{\"{u}}tt, Tkatchenko, and M{\"{u}}ller]{Unke2021}
Unke, O.~T., Chmiela, S., Sauceda, H.~E., Gastegger, M., Poltavsky, I., Sch{\"{u}}tt, K.~T., Tkatchenko, A., and M{\"{u}}ller, K.-R.
\newblock {Machine Learning Force Fields}.
\newblock \emph{Chem. Rev.}, 121:\penalty0 10142--10186, 2021.
\newblock \doi{10.1021/acs.chemrev.0c01111}.

\bibitem[van~der Oord et~al.(2023)van~der Oord, Sachs, Kov{\'a}cs, Ortner, and Cs{\'a}nyi]{vanerOord2022}
van~der Oord, C., Sachs, M., Kov{\'a}cs, D.~P., Ortner, C., and Cs{\'a}nyi, G.
\newblock Hyperactive learning for data-driven interatomic potentials.
\newblock \emph{npj Comput. Mater.}, 9\penalty0 (1):\penalty0 168, Sep 2023.
\newblock ISSN 2057-3960.
\newblock \doi{10.1038/s41524-023-01104-6}.
\newblock URL \url{https://doi.org/10.1038/s41524-023-01104-6}.

\bibitem[Vandermause et~al.(2020)Vandermause, Torrisi, Batzner, Xie, Sun, Kolpak, and Kozinsky]{Vandermause2020}
Vandermause, J., Torrisi, S.~B., Batzner, S., Xie, Y., Sun, L., Kolpak, A.~M., and Kozinsky, B.
\newblock On-the-fly active learning of interpretable bayesian force fields for atomistic rare events.
\newblock \emph{npj Comput. Mater.}, 6:\penalty0 20, 2020.

\bibitem[Verlet(1967)]{verlet1967computer}
Verlet, L.
\newblock Computer" experiments" on classical fluids. i. thermodynamical properties of lennard-jones molecules.
\newblock \emph{Physical review}, 159\penalty0 (1):\penalty0 98, 1967.

\bibitem[Yang et~al.(2022)Yang, Song, King, and Xu]{yang2022survey}
Yang, X., Song, Z., King, I., and Xu, Z.
\newblock A survey on deep semi-supervised learning.
\newblock \emph{IEEE Trans. Knowl. Data Eng.}, 2022.

\bibitem[Zaverkin \& K{\"{a}}stner(2020)Zaverkin and K{\"{a}}stner]{Zaverkin2020}
Zaverkin, V. and K{\"{a}}stner, J.
\newblock {Gaussian Moments as Physically Inspired Molecular Descriptors for Accurate and Scalable Machine Learning Potentials}.
\newblock \emph{J. Chem. Theory Comput.}, 16:\penalty0 5410--5421, 2020.
\newblock ISSN 1549-9618.
\newblock \doi{10.1021/acs.jctc.0c00347}.

\bibitem[Zaverkin \& K{\"a}stner(2021)Zaverkin and K{\"a}stner]{Zaverkin2021a}
Zaverkin, V. and K{\"a}stner, J.
\newblock Exploration of transferable and uniformly accurate neural network interatomic potentials using optimal experimental design.
\newblock \emph{Mach. Learn.: Sci. Technol.}, 2\penalty0 (3):\penalty0 035009, 2021.
\newblock \doi{10.1088/2632-2153/abe294}.
\newblock URL \url{http://iopscience.iop.org/article/10.1088/2632-2153/abe294}.

\bibitem[Zaverkin et~al.(2021)Zaverkin, Holzm{\"{u}}ller, Steinwart, and K{\"{a}}stner]{Zaverkin2021b}
Zaverkin, V., Holzm{\"{u}}ller, D., Steinwart, I., and K{\"{a}}stner, J.
\newblock {Fast and Sample-Efficient Interatomic Neural Network Potentials for Molecules and Materials Based on Gaussian Moments}.
\newblock \emph{J. Chem. Theory Comput.}, 17:\penalty0 6658--6670, 2021.
\newblock ISSN 1549-9618.
\newblock \doi{10.1021/acs.jctc.1c00527}.

\bibitem[Zaverkin et~al.(2022)Zaverkin, Holzm{\"u}ller, Steinwart, and K{\"a}stner]{Zaverkin2022d}
Zaverkin, V., Holzm{\"u}ller, D., Steinwart, I., and K{\"a}stner, J.
\newblock Exploring chemical and conformational spaces by batch mode deep active learning.
\newblock \emph{Digital Discovery}, 1:\penalty0 605--620, 2022.

\bibitem[Zaverkin et~al.(2023)Zaverkin, Holzmüller, Bonfirraro, and Kästner]{Zaverkin2023}
Zaverkin, V., Holzmüller, D., Bonfirraro, L., and Kästner, J.
\newblock Transfer learning for chemically accurate interatomic neural network potentials.
\newblock \emph{Phys. Chem. Chem. Phys.}, 25:\penalty0 5383--5396, 2023.
\newblock \doi{10.1039/D2CP05793J}.

\bibitem[Zaverkin et~al.(2024{\natexlab{a}})Zaverkin, Alesiani, Maruyama, Errica, Christiansen, Takamoto, Weber, and Niepert]{Zaverkin2024b}
Zaverkin, V., Alesiani, F., Maruyama, T., Errica, F., Christiansen, H., Takamoto, M., Weber, N., and Niepert, M.
\newblock Higher-rank irreducible cartesian tensors for equivariant message passing, 2024{\natexlab{a}}.

\bibitem[Zaverkin et~al.(2024{\natexlab{b}})Zaverkin, Holzm{\"u}ller, Christiansen, Errica, Alesiani, Takamoto, Niepert, and K{\"a}stner]{zaverkin2024uncertainty}
Zaverkin, V., Holzm{\"u}ller, D., Christiansen, H., Errica, F., Alesiani, F., Takamoto, M., Niepert, M., and K{\"a}stner, J.
\newblock Uncertainty-biased molecular dynamics for learning uniformly accurate interatomic potentials.
\newblock \emph{npj Comput. Mater.}, 10:\penalty0 83, 2024{\natexlab{b}}.

\end{thebibliography}
\bibliographystyle{icml2025}

\newpage

\appendix
\onecolumn

\setcounter{equation}{0}
\renewcommand{\theequation}{A\arabic{equation}}

\setcounter{figure}{0}
\renewcommand{\thefigure}{A\arabic{figure}}

\setcounter{table}{0}
\renewcommand{\thetable}{A\arabic{table}}

\section{Potential Broader Impact and Ethical Aspects}

This paper presents work whose goal is to advance the field of machine learning, in particular, machine learning for science. Due to the generic nature of pure science, there are many potential societal consequences of our work in the far future, none of which we feel must be specifically highlighted here.

\section{Related Work}

\textbf{Addressing Data Sparsity in MLIPs.} Generating training data sets suitable for learning reliable MLIPs is challenging, especially when considering unexplored molecular and material systems. Numerous computationally expensive calculations with either ab initio or first-principles approaches are required for the latter. To mitigate this challenge, active learning (AL) methods, which utilize prediction uncertainty, can be applied~\citep{Li2015, Podryabinkin2017, Vandermause2020, Shuaibi2021, Briganti2023,zaverkin2024uncertainty}. Furthermore, equivariant MLIPs often reduce required training data set sizes through improved data efficiency~\citep{batzner2023, batatia2022mace}.

\section{Experimental Setup, Baselines, and Data Sets}

\subsection{Experimental Setup \label{sec:setup}}

\textbf{Code for Experiments.} The code used to run our experiments builds upon the recent work~\citep{fu2023forces} and extends it to integrate the latest Open Catalyst Project code~\citep{ocp_dataset}. We adopt hyper-parameters from the Open Catalyst (OC) project, tuned to the corresponding OC data set. Note that we do not use this data set in the presented work, whose main focus is training general-purpose MLIPs that can be used to run molecular dynamics (MD) simulations and geometry optimization. However, the OC data set has been designed to investigate the latter, making it less suitable for the current study. Our modifications include adjusting the learning rate scheduler, details of which can be found in our repository. For potential energy and force prediction, we utilize mean-absolute error (MAE) and $L_2$-norm (L2MAE) losses with coefficients of 1 and 100, respectively. More details on the model hyperparameters are provided in our repository. For the PITC and PISC loss functions, we use the mean square error (MSE) loss based on an experiment in \secref{sec:neg-small-term}.

\begin{table}[b!]
    \caption{\textbf{Employed mini-batch sizes.} We provide mini-batch sized for all data sets and models employed in this work. \label{tb:mini_batch_size}}
    \begin{center}
    \begin{tabular}{ccccc}
    \toprule
                    & ANI-1x & TiO$_2$\textsuperscript{\emph{a}}    & rMD17 & LMNTO     \\
    \cmidrule(lr){2-5}
    Mini-batch size & 6      & 4                                    & 16    & 4         \\
    \bottomrule
    \end{tabular}
    \end{center}
    \footnotesize{\textsuperscript{\emph{a}} As explained in \tabref{tab:tio2-results}, the mini-batch size of the SchNet model was changed to 32 due to high RMSE values observed with a mini-batch size of four as the training data set size increased. A more detailed discussion of the results for SchNet is provided in \secref{sec:add-benchmark}.}
\end{table}

\begin{table}[t!]
    \caption{\textbf{Total number of training iterations.} We provide the total number of training iterations for all data sets and training set sizes employed in this work. The number in the parentheses demonstrates the corresponding total number of training epochs. \label{tb:iteration_num}}
    \centering
    \begin{tabular}{rrrrr}
    \toprule
    $N_\mathrm{train}$   & ANI-1x                    & TiO2              & rMD17                         & LMNTO             \\
    \cmidrule(lr){1-1} \cmidrule(lr){2-5}
    50                   & 7500 (900)                & --                & --                            & --                \\
    100                  & 10,000 (600)              & 10,000 (400)      & 7500 (1200)                   & 10,000 (400)      \\
    1000                 & 40,000 (240)              & 10,000 (100)      & 10,000 (160)                  & 10,000 (100)      \\
    10,000               & 100,000 (60)              & --                & --                            & --                \\
    100,000              & 400,000 (30)              & --                & --                            & --                \\
    1,000,000            & 400,000 (3)              & --                & --                            & --                \\
    5,000,000            & 420,000 (1)              & --                & --                            & --                \\
    \bottomrule
    \end{tabular}
\end{table}

\textbf{Training Details.} For training MLIPs, we followed the setup in the Open Catalyst Project. We kept the mini-batch size consistent across all models, as shown in \tabref{tb:mini_batch_size}. We have chosen the mini-batch size based on the maximum memory needed by the most demanding models, such as eSCN and Equiformer v2. All experiments are performed on a single NVIDIA A100 GPU with 81.92 GB memory. To avoid overfitting, we stopped training when the validation loss stopped improving---the specific number of training iterations is provided in \tabref{tb:iteration_num}.

We used perturbation vectors $\delta \mathbf{r}$ drawn from a uniform random distribution; see also \secref{sec:sub-perturbation-types}. Particularly, we defined $\delta \mathbf{r} \equiv \epsilon \mathbf{g}$, with each component of $\mathbf{g}$ drawn from a uniform random distribution in the interval $(-1, 1)$. The magnitude $\epsilon$ is also drawn from a uniform random distribution and $\epsilon < \epsilon_\mathrm{max}$. This definition of $\delta \mathbf{r}$ differs from the one in \eqref{eq:adv-random}, improving the computational efficiency of PIWSL by avoiding the calculation of square root and division.

The remaining hyper-parameters are the coefficients for the PITC and PISC losses ($C_{\rm PITC}, C_{\rm PISC}$) and the maximum magnitude $\epsilon_\mathrm{max}$ of the perturbation vector $\delta \mathbf{r}$; see \tabref{tb:hyper-parameter-list}. These hyper-parameters are tuned using Optuna~\citep{optuna_2019} for PaiNN and Equiformer v2. We used 1000 configurations drawn randomly from the original ANI-1x data set for training. Due to multiple local minima, Optuna identified several optimal hyper-parameter sets in each run. We selected the following representative combinations ($C_{\rm PITC}, C_{\rm PISC}, \epsilon_\mathrm{max})$ = Case A: (1.2, 0.8, 0.025), Case B: (1.0, 0, 0.01), Case C: (0.1, 0.01, 0.01), Case D: (1.2, 0.01, 0.025), Case E: (1.2, 0.01, 0.01), Case F: (1.2, 0.01, 0.015), Case G: (0.01, 0.001, 0.025), and Case H: (0.1, 0.01, 0.025). We selected the hyper-parameters listed in \tabref{tb:hyper-parameter-list} based on the validation dataset performance. 

\begin{table*}
    \caption{\textbf{Hyper-parameters for the PIWSL loss.} We selected the following hyper-parameter combinations using Optuna~\citep{optuna_2019}: ($C_{\rm PITC}, C_{\rm PISC}, \epsilon_\mathrm{max})$= Case A: (1.2, 0.8, 0.025), Case B: (1.0, 0, 0.01), Case C: (0.1, 0.01, 0.01), Case D: (1.2, 0.01, 0.025), Case E: (1.2, 0.01, 0.01), Case F: (1.2, 0.01, 0.015), Case G: (0.01, 0.001, 0.025), and Case H: (0.1, 0.01, 0.025). \label{tb:hyper-parameter-list}}
    \centering
    \begin{tabular}{lrccccc}
    \toprule
    Dataset         & Size      & Equiformer v2 & eSCN  & PaiNN     & SpinConv  & SchNet    \\
    \cmidrule(lr){1-2} \cmidrule(lr){3-7}
    \multirow[c]{4}{*}{ANI-1x}                & 50        & A             & C     & B         & A         & A         \\
              & 100       & A             & C     & A         & D         & A         \\
                    & 1000      & D             & D     & D         & B         & B         \\
                    & 10,000    & G             & C     & B         & C         & C         \\
                    & 100,000    & --             & --     & C         & --        & --         \\
                    & 1,000,000    & --            & --     & C         & --         & --         \\
                    & 5,000,000    & --            & --     & B     & --         & --         \\
    \cmidrule(lr){1-2} \cmidrule(lr){3-7}
    \multirow[c]{2}{*}{TiO2}            & 100       & A             & A     & A         & A         & H         \\
                    & 1000      & G             & A     & C         & A         & C         \\
    \cmidrule(lr){1-2} \cmidrule(lr){3-7}
    \multirow[c]{2}{*}{LMNTO}           & 100       & B             & B     & B         & A         & B         \\
                    & 1000      & B             & A     & B         & B         & B         \\
    \bottomrule
    \end{tabular}
\end{table*}

\textbf{Splitting Data Sets.} We split the original data sets into training, validation, and test sets for our experiments. We shuffled the original data sets using a random seed and selected the training data sets of predefined sizes. For validation, we selected the same number of configurations as in the training data set if it exceeded 100 configurations; otherwise, we used 100 configurations to ensure sufficient validation size. For the rMD17 data set, following \citep{fu2023forces}, we used 9000 configurations as a validation data set and another 10,000 for testing. We used the same test data set across different sizes of the training data sets for a fair performance comparison. We used 10,000 test configurations for ANI-1x and 1000 for TiO$_2$ and LMNTO.

\textbf{Training with Adversarial Directions.} In our experiments, which defined perturbation vectors adversarially (see \secref{sec:analysis}), we determined adversarial directions using \eqref{eq:adv-direc}. More concretely, we only considered the potential energy, i.e., $\mathbf{y}_{\rm pred}$ and $\mathbf{y}_{\rm label}$, to avoid Hessian calculations. In addition, we considered the loss function for the potential energy as $L_{\rm dist}$ in \eqref{eq:adv-direc}. The expression of $\mathbf{g} = \nabla_{\mathbf r} L_{\rm dist}$ is then
\begin{equation}
    \mathbf{g}_{\mathcal{S}} = \nabla_{\mathbf{r}_i} \sqrt{(E \left(\mathcal{S}; \boldsymbol{\theta}\right) - E^{\rm ref}_{\mathcal{S}})^2} = - \frac{1}{2 L_{\rm dist}}(\mathbf{F}_i\left(\mathcal{S}; \boldsymbol{\theta}\right) - \mathbf{F}^{\rm ref}_{i,\mathcal{S}}) (E\left(\mathcal{S}; \boldsymbol{\theta}\right) - E^{\rm ref}_{\mathcal{S}}). 
\end{equation}
Note that we used the relation $\nabla_{\mathbf{r}_i} E^{\rm ref}_{\mathcal{S}} = - \mathbf{F}^{\rm ref}_{i,\mathcal{S}}$ to obtain the final expression. Though $E^{\rm ref}$ can also be interpreted as a constant regarding atom positions. We have chosen this expression to avoid the case where the adversarial direction $\mathbf{g}$ points to $\mathbf{F}^{\rm ref}_{i,\mathcal{S}}$ other than the very beginning of the training. Our experiments indicate that the employed expression is slightly better than its alternative. In our experiment, we also randomly flip the sign of $\mathbf{g}_{\mathcal{S}}$ to avoid overfitting to adversarial directions.

\subsection{Baseline Methods}

\textbf{Data Augmentation with NoisyNode.} In our experiments, we used the NoisyNode approach~\citep{godwin2021simple} as one of the baseline methods. This method aims to improve the performance of ML models by adding a perturbation to node features, i.e., atomic coordinates, and makes ML models recover original labels. This approach enables ML models to be more robust to noise in the data. Although the original method recommends adding a decoder network to learn the denoising process, we do not utilize it following previous work~\citep{equiformer_v2} and add the perturbation vector to atomic coordinates similar to PIWSL losses, fixing energy and force labels. We implement the NoisyNode approach in our code. Thus, we can expect slightly different behavior compared to the recent work~\citep{godwin2021simple,equiformer_v2}~\footnote{Note that NoisyNode assumes the unperturbed state to be the equilibrium structure, which may have contributed to the limited performance improvements observed in our experiments. Recently, efforts have been underway to develop an extended version of NoisyNode tailored for non-equilibrium structures, such as \citep{DeNS}.}.

\textbf{Taylor-Expansion-Based Weak Labels.} Recent work proposed a similar Taylor-expansion-based weak label approach~\citep{cooper2020efficient}. Nonetheless, the loss is different from the one in \eqref{eq:taylor-wsl-loss} as the authors used reference energy and atomic force labels to estimate weak energy labels $E^\mathrm{ref}_{\mathcal{S}_{\dr}}$ for perturbed atomic configurations $\mathcal{S}_{\dr}$
\begin{equation}
    \label{eq:taylor-nong}
    \begin{split}
        E^\mathrm{ref}_{\mathcal{S}_{\dr}} \approx E^\mathrm{ref}_{\mathcal{S}} - \sum_{i=1}^{N_\mathrm{at}} \left\langle \dr_i, \mathbf{F}^\mathrm{ref}_{i,\mathcal{S}} \right\rangle + \mathcal{O}\left(||\dr||^2\right).
    \end{split}
\end{equation}
The trainable parameters of MLIPs are optimized by minimizing the weak label (WL) loss
\begin{equation}
    \label{eq:taylor-nong-loss}
    \begin{split}
        L_\mathrm{WL}\left(\mathcal{S};\boldsymbol{\theta}\right) &  = \ell\left( E\left(\mathcal{S}_{\dr}; \boldsymbol{\theta}\right), E^\mathrm{ref}_{\mathcal{S}} - \sum_{i=1}^{N_\mathrm{at}} \left\langle \dr_i, \mathbf{F}^\mathrm{ref}_{i,\mathcal{S}} \right\rangle \right).
    \end{split}
\end{equation}

\Figref{fig:weakly-supervised} (a) illustrates the corresponding approach~\citep{cooper2020efficient}, which computes the energy of a perturbed atomic configuration using a Taylor expansion based on reference energy and atomic force labels. This approach was originally applied to train MLIPs without explicit force labels.

\subsection{Description of the Data Sets \label{sec:dataset-explanation}}

\textbf{ANI-1x Data Set.} The ANI-1x data set is a heterogeneous molecular data set and includes 63,865 organic molecules (with chemical elements H, C, N, and O) whose size ranges from 4 to 64 atoms~\citep{smith2020ani}. The ML model requires learning total energies and atomic forces for various molecules and their conformations. Total energies and atomic forces are obtained through DFT calculations.

\textbf{TiO$_2$ Data Set.} Titanium dioxide (TiO$_2$) is an industrially relevant and well-studied material. TiO$_2$ dataset includes 7815 bulk structures of several TiO$_2$ phases whose reference energies and forces are obtained through DFT calculations~\citep{ARTRITH2016135}. The number of atoms in a single configuration ranges from 6 to 95.

\textbf{rMD17 Data Set.} The rMD17 data set includes ten small organic molecules, including 100,000 configurations obtained by running MD simulations for each~\citep{Christensen2020}. The ML model requires learning the total energies and atomic forces for each molecule. In this revised version of the MD17 data set, the molecules are taken from the original MD17 data set~\citep{chmiela2017machine,chmiela2018towards}. However, the energies and forces are recalculated at the PBE/def2-SVP level of theory using very tight SCF convergence and a very dense DFT integration grid.

\textbf{LMNTO Data Set.} The Li-Mo-Ni-Ti oxide (LMNTO) is of technological significance as a potential high-capacity positive electrode material for lithium-ion batteries. It exhibits substitutional disorder, with Li, Mo, Ni, and Ti all sharing the same sublattice. This data set includes LMNTO with the composition  Li$_8$Mo$_2$Ni$_7$Ti$_7$O$_{32}$ and configurations obtained from an MD simulation, resulting in approximately 2600 structures in total~\citep{cooper2020efficient}.

\textbf{MD22 Data Set.} The MD22 data set ~\citep{md22} includes seven larger organic molecules, such as a small peptide and a double-walled nanotube, whose size ranges from 42 to 370 atoms. The data set consists of MD trajectories sampled at temperatures between 400 and 500 K. The ML model requires learning the total energies and atomic forces for each molecule. The energies and forces are calculated at the PBE+MBD level of theory.

\section{Differences in Gradients for Physics-Informed Losses \label{sec:loss-diff}}

The following considers the gradients of the proposed two losses. First, considering squared errors, we obtain the following gradients of the loss in \eqref{eq:taylor-nong} with respect to trainable parameters
\begin{equation}
    \begin{split}
    & \frac{\mathrm{d}\mathcal{L}_\mathrm{WL}}{\mathrm{d}\boldsymbol{\theta}} = 2 \Bigg( E\left(\mathcal{S}_{\dr}; \boldsymbol{\theta}\right) - E^\mathrm{ref}_{\mathcal{S}} + \sum_{i=1}^{N_\mathrm{at}} \langle \dr_i, \mathbf{F}^\mathrm{ref}_{i,\mathcal{S}} \rangle\Bigg) \frac{\mathrm{d}E\left(\mathcal{S}_{\dr}; \boldsymbol{\theta}\right)}{\mathrm{d}\boldsymbol{\theta}}.
    \end{split}
\end{equation}
In contrast, for the PITC loss in \eqref{eq:taylor-wsl-loss} we obtain
\begin{equation}
    \label{eq:dif-taylor-wsl-loss}
    \begin{split}
    \frac{\mathrm{d}\mathcal{L}_\mathrm{PITC}}{\mathrm{d}\boldsymbol{\theta}} = & 2 \left(E\left(\mathcal{S}_{\dr}; \boldsymbol{\theta}\right) - E\left(\mathcal{S}; \boldsymbol{\theta}\right) + \sum_{i=1}^{N_\mathrm{at}} \langle \dr_i, \mathbf{F}_i\left(\mathcal{S}; \boldsymbol{\theta}\right) \rangle\right) \times \\ & \left(\frac{\mathrm{d}E\left(\mathcal{S}_{\dr}; \boldsymbol{\theta}\right)}{\mathrm{d}\boldsymbol{\theta}} - \frac{\mathrm{d}E\left(\mathcal{S}; \boldsymbol{\theta}\right)}{\mathrm{d}\boldsymbol{\theta}} + \sum_{i=1}^{N_\mathrm{at}} \frac{\mathrm{d}\langle \dr_i, \mathbf{F}_i\left(\mathcal{S}; \boldsymbol{\theta}\right) \rangle}{\mathrm{d}\boldsymbol{\theta}}\right).
    \end{split}
\end{equation}
The above equations indicate that the direction of the derivative of the PITC loss in \eqref{eq:dif-taylor-wsl-loss} is different from that of the weak label loss because of the incorporation of the predicted potential energy at the original and the force at the reference point. The gradient of PISC loss in \eqref{eq:spatial-consis-loss} reads
\begin{equation}
    \label{eq:dif-sp-wsl-loss}
    \begin{split}
    \frac{\mathrm{d}\mathcal{L}_\mathrm{PISC}}{\mathrm{d}\boldsymbol{\theta}} &= 2 \Bigg( E\left(\mathcal{S}; \boldsymbol{\theta}\right) - \sum_{i=1}^{N_\mathrm{at}} \langle \dr_i, \mathbf{F}_i\left(\mathcal{S}; \boldsymbol{\theta}\right) \rangle - E\left(\mathcal{S}_{\dr^\prime}; \boldsymbol{\theta}\right) + \sum_{i=1}^{N_\mathrm{at}} \langle \dr_i^{\prime\prime}, \mathbf{F}_i\left(\mathcal{S}_{\dr^\prime}; \boldsymbol{\theta}\right) \rangle\Bigg) \\ & \times \Bigg( \frac{\mathrm{d}E\left(\mathcal{S}; \boldsymbol{\theta}\right)}{\mathrm{d}\boldsymbol{\theta}} - \sum_{i=1}^{N_\mathrm{at}} \frac{\mathrm{d}\langle \dr_i, \mathbf{F}_i\left(\mathcal{S}; \boldsymbol{\theta}\right) \rangle}{\mathrm{d}\boldsymbol{\theta}} - \frac{\mathrm{d}E\left(\mathcal{S}_{\dr^\prime}; \boldsymbol{\theta}\right)}{\mathrm{d}\boldsymbol{\theta}} + \sum_{i=1}^{N_\mathrm{at}} \frac{\mathrm{d}\langle \dr_i^{\prime\prime}, \mathbf{F}_i\left(\mathcal{S}_{\dr^\prime}; \boldsymbol{\theta}\right) \rangle}{\mathrm{d}\boldsymbol{\theta}}\Bigg).
    \end{split}
\end{equation}

\section{Experiments}

\subsection{Benchmark Results \label{sec:add-benchmark}}

The following section provides additional results, complementing those provided in the main text. 

\begin{table*}[t!]
    \caption{\textbf{Energy (E) and force (F) root-mean-square errors (RMSEs) for the ANI-1x data set.} The results are obtained by averaging over three independent runs. Energy RMSE is given in kcal/mol, while force RMSE is in kcal/mol/\AA. \label{tab:ani-1x-results-5010K}}
    \begin{center}
    \resizebox{\columnwidth}{!}{
    \begin{tabular}{llcccccc}
    \toprule
    & & \multicolumn{3}{c}{$N_\mathrm{train} = 50$}  &  \multicolumn{3}{c}{$N_\mathrm{train} = 10,000$}  \\
                                &           & Baseline                    & Noisy Nodes               & PIWSL                         & Baseline                        & Noisy Nodes                   & PIWSL                         \\
    \cmidrule(lr){1-2} \cmidrule(lr){3-5} \cmidrule(lr){6-8}
    \multirow[c]{2}{*}{Schnet}      & E    & 90.08 $\pm$ 1.24      & {\bf 76.83 $\pm$ 0.75}    & 83.90 $\pm$ 2.82              & 24.88 $\pm$ 0.01          & {\bf 24.86 $\pm$ 0.00}        & 24.88 $\pm$ 0.00              \\
                                & F     & 35.49 $\pm$ 0.36   & {\bf 31.13 $\pm$ 0.13}       & 35.30 $\pm$ 0.87              & 13.36 $\pm$ 0.01          & 13.36 $\pm$ 0.00        & 13.36 $\pm$ 0.00              \\
    \cmidrule(lr){1-2} \cmidrule(lr){3-5} \cmidrule(lr){6-8}
    \multirow[c]{2}{*}{PaiNN}       & E    & 212.64 $\pm$ 1.14  & 440.11 $\pm$ 11.68            & {\bf 121.36 $\pm$ 4.13}      & 19.14 $\pm$ 0.38          & 165.25 $\pm$ 4.87            & {\bf 14.10 $\pm$ 0.14}        \\  
                                & F     & 22.61 $\pm$ 0.04   & 22.50 $\pm$ 0.22             & {\bf 20.83 $\pm$ 0.28}        & 8.24 $\pm$ 0.10          & 9.22 $\pm$ 0.09              & {\bf 7.89 $\pm$ 0.02}        \\
    \cmidrule(lr){1-2} \cmidrule(lr){3-5} \cmidrule(lr){6-8}
    \multirow[c]{2}{*}{SpinConv}    & E    & 222.75 $\pm$ 7.12  & 219.85 $\pm$ 6.99            & {\bf 175.38 $\pm$ 9.77}       & 19.42 $\pm$ 0.67          & 46.31 $\pm$ 10.31           & {\bf 18.81 $\pm$ 0.60}        \\
                                & F     & {\bf 24.88 $\pm$ 0.88}   & 24.61 $\pm$ 0.35       & 25.12 $\pm$ 0.58              & 10.31 $\pm$ 0.33          & 10.78 $\pm$ 0.66              & {\bf 9.94 $\pm$ 0.12}        \\
    \cmidrule(lr){1-2} \cmidrule(lr){3-5} \cmidrule(lr){6-8}
    \multirow[c]{2}{*}{eSCN}        & E    & 517.17 $\pm$ 31.98  & 583.90 $\pm$ 33.04           & {\bf 454.40 $\pm$ 11.10}       & 12.65 $\pm$ 0.63          & 165.30 $\pm$ 33.11            & {\bf 10.66 $\pm$ 0.31}        \\
                                & F     & 22.51 $\pm$ 0.09   & 24.04 $\pm$ 0.15             & {\bf 22.28  $\pm$ 0.08}       & 5.11 $\pm$ 0.30          & 11.51 $\pm$ 0.23              & {\bf 4.35 $\pm$ 0.15}        \\
    \cmidrule(lr){1-2} \cmidrule(lr){3-5} \cmidrule(lr){6-8}
    \multirow[c]{2}{*}{Equiformer}  & E    & 498.58 $\pm$ 17.44 & 630.32 $\pm$ 0.32            & {\bf 433.88 $\pm$ 79.63}       & 8.03 $\pm$ 0.21          & 970.95 $\pm$ 236.90            & {\bf 7.77 $\pm$ 0.14}        \\
                                & F     & 22.86 $\pm$ 0.04   & 22.92 $\pm$ 0.00             & {\bf 22.72 $\pm$ 0.04}        & {\bf 2.97 $\pm$ 0.00}          & 29.28 $\pm$ 5.63              & 2.98 $\pm$ 0.00         \\
    \bottomrule
    \end{tabular}}
    \end{center}
\end{table*}

\begin{table}[h]
    \caption{\textbf{Energy and force erorrs of PaiNN model trained on the ANI-1x data set with 100,000, 1,000,000, and 5,000,000 samples.} The results are obtained by averaging over three independent runs. Energy errors are given in kcal/mol, while force errors are in kcal/mol/\AA. \label{tb:ani-1x-100K}}
    \begin{center}
    \resizebox{0.8\columnwidth}{!}{
    \begin{tabular}{crrrrr}
    \toprule
                & $N_{\rm train}$ & Force MAE & Force RMSE  & Energy MAE & Energy RMSE \\
    \cmidrule(lr){1-6}
    Baseline    & 100,000 & {\bf 0.92 $\pm$ 0.00} & {\bf 3.70 $\pm$ 0.01} & 4.28 $\pm$ 0.15 & 6.14 $\pm$ 0.21 \\
    PIWSL       & & {\bf 0.91 $\pm$ 0.01} & {\bf 3.72 $\pm$ 0.04} & {\bf 4.07 $\pm$ 0.17} & {\bf 5.83 $\pm$ 0.20} \\
    \midrule
    Baseline    & 1,000,000 & {\bf 0.67 $\pm$ 0.00} & {\bf 2.74 $\pm$ 0.01} & 4.90 $\pm$ 0.23 & 6.56 $\pm$ 0.26 \\
    PIWSL       & & 0.68 $\pm$ 0.00 & {\bf 2.77 $\pm$ 0.04} & {\bf 4.48 $\pm$ 0.05} & {\bf 6.06 $\pm$ 0.01} \\
    \midrule
    Baseline    & 5,000,000\textsuperscript{\emph{a}} & {\bf 0.53} & {\bf 2.18} & 3.94 & 5.24 \\
    PIWSL       & & 0.55 & 2.24 & {\bf 3.25} & {\bf 4.75} \\
    \bottomrule
    \end{tabular}}
    \end{center}    
    \footnotesize{\textsuperscript{\emph{a}} Because of the computational cost, we performed only one training in the case of 5,000,000 training samples.} \\
\end{table}

\textbf{Additional Results for ANI-1x.} \Tabref{tab:ani-1x-results-5010K} provides results for ANI-1x data set and a training set sizes of 50 or 10,000; see also \figref{fig:enter-label}. The table demonstrates a considerable reduction of energy and force RMSEs for models trained using small training data set sizes of 50 configurations. Furthermore, we find around 5 to 25~\% error reduction for a larger training set size of 10,000, indicating the effectiveness of the PIWSL method for relatively large training set sizes. 
Finally, we provide a result of PaiNN model trained on ANI-1x data set with 100,000, 1,000,000, and 5,000,000 samples in \Tabref{tb:ani-1x-100K}, which demonstrates that PIWSL still improves the performance around 5\%  to 10\%of the energy RMSE.

\begin{table}[t]
    \caption{\textbf{Energy (F) and force (F) root-mean-square errors (RMSEs) for the TiO2 data set obtained for the SchNet model with a mini-batch size of four.} The results are obtained by averaging over three independent runs. Energy RMSE is given in kcal/mol, while force RMSE is in kcal/mol/\AA. \label{tb:SchNet-TiO2-bt4}}
    \begin{center}
    \resizebox{0.95\columnwidth}{!}{
    \begin{tabular}{lcrrrrrr}
    \toprule
                                &   & \multicolumn{3}{c}{$N_\mathrm{train} = 100$}                                        &  \multicolumn{3}{c}{$N_\mathrm{train} = 1000$}                                      \\
                                &   & Baseline          & Noisy Nodes                    & PIWSL                          & Baseline          & Noisy Nodes                    & PIWSL                          \\
    \cmidrule(lr){1-2} \cmidrule(lr){3-5} \cmidrule(lr){6-8}
    \multirow[c]{2}{*}{SchNet}  & E & 18.85 $\pm$ 0.00  & {\bf 17.48 $\pm$ 0.00}         & 17.58 $\pm$ 0.00               & 35.58 $\pm$ 0.00  & 58.08 $\pm$ 18.44              & {\bf 15.28 $\pm$ 0.12}         \\
                                & F & 2.74 $\pm$ 0.00   & {\bf 2.51 $\pm$ 0.00}          & 2.74 $\pm$ 0.00                & 6.54 $\pm$ 0.00   & 18.40 $\pm$ 0.00               & {\bf 3.61 $\pm$ 0.27}          \\
    \bottomrule
    \end{tabular}}
    \end{center}
\end{table}

\begin{table}[h]
    \caption{\textbf{Training data set size dependence of SchNet with a mini-batch size of four.} The results are presented for the TiO$_2$ data set and are obtained by averaging over three independent runs. Energy RMSE is given in kcal/mol, while force RMSE is in kcal/mol/\AA. \label{tb:SchNet-TiO2-Sample}}
    \begin{center}
    \begin{tabular}{llrrrr}
    \toprule
                               &    & $N_\mathrm{train} = 100$  & $N_\mathrm{train} = 200$  & $N_\mathrm{train} = 500$  & $N_\mathrm{train} = 1000$     \\    
    \cmidrule(lr){1-2} \cmidrule(lr){3-6}
    \multirow[c]{2}{*}{SchNet} & E  & 18.85 $\pm$ 0.00          & {\bf 16.28 $\pm$ 0.00}    & 24.42 $\pm$ 0.00          & 35.58 $\pm$ 0.00              \\
                               & F  & 2.74 $\pm$ 0.00           & {\bf 2.56 $\pm$ 0.00}     & 4.43 $\pm$ 0.00           & 6.54 $\pm$ 0.00               \\
    \bottomrule
    \end{tabular}
    \end{center}
\end{table}

\textbf{Additional Results for SchNet Applied to TiO$_2$.} In \tabref{tab:tio2-results}, we set the mini-batch size to 32 for training the SchNet model. This adjustment was made because training SchNet with a small mini-batch size of four increases RMSE values with a growing training data set size. \Tabref{tb:SchNet-TiO2-bt4} demonstrates the performance of the SchNet model for a mini-batch size of four. \Tabref{tb:SchNet-TiO2-Sample} provides the results obtained for the SchNet model with a mini-batch size of four for the following training set sizes: 100, 200, 500, and 1000. This figure demonstrates that SchNet, with a mini-batch size of four, reaches its best performed with $N_{\rm train} = 200$. These results indicate the difficulty of learning training data statistics from small mini-batches, probably due to the limited expressive power of SchNet.

\begin{table*}
    \begin{center}
    \caption{\textbf{Energy (E) and force (F) root-mean-square errors (RMSEs) for the LMNTO data set.} The results are obtained by averaging over three independent runs. Energy RMSE is given in kcal/mol, while force RMSE is in kcal/mol/\AA. \label{tb:LMNTO_results}}
    \resizebox{0.9\columnwidth}{!}{
    \begin{tabular}{llrrrrrr}
    \toprule
                                        &   & \multicolumn{3}{c}{$N_\mathrm{train} = 100$}                                  &  \multicolumn{3}{c}{$N_\mathrm{train} = 1000$}                    \\
    Model                               &   & Baseline              & NoisyNode                 & PIWSL                     & Baseline              & NoisyNode         & PIWSL                 \\
    \cmidrule(lr){1-2} \cmidrule(lr){3-5} \cmidrule(lr){6-8}
    \multirow[c]{2}{*}{SchNet}          & E & 4.46 $\pm$ 0.00       & 6.10 $\pm$ 0.00           & {\bf 4.45 $\pm$ 0.00}     & {\bf 3.09 $\pm$ 0.00} & 3.25 $\pm$ 0.00   & {\bf 3.09 $\pm$ 0.00} \\
                                        & F & 9.24 $\pm$ 0.00       & {\bf 8.31 $\pm$ 0.00}     & 9.24 $\pm$ 0.00           & {\bf 5.09 $\pm$ 0.00} & 5.21 $\pm$ 0.00   & {\bf 5.09 $\pm$ 0.00} \\
    \cmidrule(lr){1-2} \cmidrule(lr){3-5} \cmidrule(lr){6-8}
    \multirow[c]{2}{*}{PaiNN}           & E & 6.91 $\pm$ 0.02       & 7.09 $\pm$ 0.04           & {\bf 5.99 $\pm$ 0.02}     & 3.26 $\pm$ 0.01       & 4.61 $\pm$ 0.03   & {\bf 2.98 $\pm$ 0.01} \\
                                        & F & {\bf 4.75 $\pm$ 0.00} & 7.20 $\pm$ 0.01           & {\bf 4.75 $\pm$ 0.00}     & {\bf 2.03 $\pm$ 0.00} & 2.55 $\pm$ 0.00   & {\bf 2.03 $\pm$ 0.00} \\
    \cmidrule(lr){1-2} \cmidrule(lr){3-5} \cmidrule(lr){6-8}
    \multirow[c]{2}{*}{SpinConv}        & E & 7.90 $\pm$ 0.00       & 7.83 $\pm$ 0.04           & {\bf 7.83 $\pm$ 0.01}     & 4.90 $\pm$ 0.33       & 7.20 $\pm$ 0.06   & {\bf 3.95 $\pm$ 0.02} \\
                                        & F & {\bf 4.63 $\pm$ 0.01} & 5.14 $\pm$ 0.04           & 4.71 $\pm$ 0.02           & 1.81 $\pm$ 0.01       & 2.33 $\pm$ 0.00   & {\bf 1.74 $\pm$ 0.00} \\
    \cmidrule(lr){1-2} \cmidrule(lr){3-5} \cmidrule(lr){6-8}
    \multirow[c]{2}{*}{eSCN}            & E & 7.92 $\pm$ 0.00       & 7.92 $\pm$ 0.00           & 7.92 $\pm$ 0.00           & 7.93 $\pm$ 0.00       & 7.93 $\pm$ 0.00   & {\bf 6.40 $\pm$ 0.14} \\
                                        & F & 4.67 $\pm$ 0.01       & 7.59 $\pm$ 0.02           & {\bf 4.64 $\pm$ 0.01}     & 1.54 $\pm$ 0.00       & 1.98 $\pm$ 0.06   & {\bf 1.53 $\pm$ 0.00} \\
    \cmidrule(lr){1-2} \cmidrule(lr){3-5} \cmidrule(lr){6-8}
    \multirow[c]{2}{*}{Equiformer v2}   & E & 7.40 $\pm$ 0.03       & 7.92 $\pm$ 0.00           & {\bf 7.32 $\pm$ 0.08}     & {\bf 3.57 $\pm$ 0.05} & 7.04 $\pm$ 0.03   & 3.60 $\pm$ 0.02       \\
                                        & F & 4.26 $\pm$ 0.00       & 7.60 $\pm$ 0.02           & {\bf 4.24 $\pm$ 0.02}     & {\bf 1.34 $\pm$ 0.00} & 1.99 $\pm$ 0.00   & {\bf 1.34 $\pm$ 0.00} \\
    \bottomrule
    \end{tabular}}
    \end{center}
\end{table*}

\textbf{Results for LMNTO.} \Tabref{tb:LMNTO_results} presents RMSE errors for LMNTO~\citep{cooper2020efficient}. PIWSL shows the error reduction for most cases for this benchmark data set, especially for small training set sizes (i.e., a training set of 100 configurations).

\begin{table*}
    \begin{center}
    \caption{\textbf{Energy (E) and force (F) mean-absolute errors (MAEs) for the MD22 data set.} We have chosen buckyball catcher for our experiments. The results are obtained by averaging over three independent runs. Energy MAE is given in kcal/mol/atom, while force MAE is in kcal/mol/\AA.}
    \label{tb:MD22_buckyball_results}
    \resizebox{0.9\columnwidth}{!}{
    \begin{tabular}{lllrrrr}
    \toprule
                        &                                   &   & \multicolumn{2}{c}{$N_\mathrm{train} = 50$}                              &  \multicolumn{2}{c}{$N_\mathrm{train} = 600$}                            \\
    Dataset             & Model                             &   & Baseline              & PIWSL                 & Baseline              & PIWSL                     \\
    \cmidrule(lr){1-1} \cmidrule(lr){2-3} \cmidrule(lr){4-5} \cmidrule(lr){6-7}
    Buckyball catcher   & \multirow[c]{2}{*}{MACE}  & E & 1.046 $\pm$ 0.095  & {\bf 0.745 $\pm$ 0.031}  & {\bf 0.587 $\pm$ 0.107} & {\bf 0.548 $\pm$ 0.015}     \\
                        &                           & F & 0.294 $\pm$ 0.002 & {\bf 0.290 $\pm$ 0.002} & {\bf 0.082 $\pm$ 0.001} & {\bf 0.082 $\pm$ 0.001}     \\
    \bottomrule
    \end{tabular}}
    \end{center}
\end{table*}

\textbf{Molecular Dynamics Trajectories for Large Molecules -- MD22.} \Tabref{tb:MD22_buckyball_results} evaluates the impact of PIWSL on data sets containing conformations of a single large molecule. For this purpose, we selected the buckyball catcher molecule with $n_{\rm atom} = 148$ atoms. To demonstrate the applicability of PIWSL, we trained a MACE model \citep{batatia2022mace} to prove that PIWSL enhances even the performance of a recent state-of-the-art model. The model structure and training configuration were based on those provided in the official repository\footnote{\url{https://mace-docs.readthedocs.io/en/latest/examples/training_examples.html}} with slight modifications: ``max\_$L$'' was set to one and mini-batch size was adjusted to four. Following the original training setup \citep{md22}, we randomly sampled 600 configurations for the training dataset and 400 for the validation dataset and retained the remaining 5102 configurations for testing. To further validate PIWSL's effectiveness in sparse data scenarios, we prepared a smaller training dataset comprising only 50 configurations while keeping the validation dataset size unchanged. The model was trained for 450 and 800 epochs for the 600-sample and 50-sample training datasets, respectively. 
The results presented in \tabref{tb:MD22_buckyball_results} demonstrate that our approach remains effective on average, particularly in the sparse data regime. In this study, the coefficient of the PITC and PISC losses are set as 0.01 and 0.001 with $\epsilon_{\rm max} = 0.01$.

In addition to the experiment of training from scratch, we conducted an additional study using the MACE foundation model to evaluate whether PIWSL could be effectively applied in the context of foundation models with finetuning. Specifically, we utilized two foundation models: MACE-MP ("large") \citep{Batatia2023}, a universal model, and MACE-OFF ("large\_off") \citep{Kovacs2023}, a model designed for organic force fields. Training was performed on the buckyball catcher molecule from the MD22 dataset. To simulate finetuning on a smaller dataset, the models were trained using 50 samples, consistent with the previous experiment. The official settings from the respective repositories were used, with adjustments made to the mini-batch size (set to 4) and the number of epochs (set to 100).

The results, presented in \Tabref{tb:MD22_buckyball_results_FM}, demonstrate that PIWSL is effective for the fine-tuning of foundation models. Moreover, the result indicates that the MACE-MP model provides worse results than MACE-OFF, emphasizing the importance of selecting an appropriate pre-trained model. For this experiment, we employed the second-order PITC loss and the two-point consistency loss defined in \eqref{eq:taylor-2nd-wsl-loss} and \eqref{eq:spatial-2pt-consis-loss}, with the coefficients set to 0.08.

\subsection{Training Time Analysis}

\begin{table}[h]
    \caption{\textbf{Training time comparison for experiments with and without PIWSL.} We measure the time required for a single training epoch and provide the results obtained as an average over five epochs. We use 1000 configurations from the ANI-1x data set. All training times are provided in seconds. \label{tb:training_time}}
    \centering
    \begin{tabular}{crrrrr}
    \toprule
                & SchNet    & PaiNN     & SpinConv  & eSCN      & Equiformer v2 \\
    \cmidrule(lr){1-1} \cmidrule(lr){2-6}
    Baseline    & 7.51      & 8.02      & 33.46     & 100.71    & 57.79         \\
    PIWSL       & 12.84     & 23.48     & 86.28     & 328.48    & 177.55        \\
    \bottomrule
    \end{tabular}
\end{table}

\begin{table}[h]
    \caption{\textbf{Training time comparison for experiments with and without PIWSL on MACE-OFF.} We measure the time required for a single training epoch and provide the results obtained as an average over five epochs. We use 950 configurations from the MD17-CC data set. All training times are provided in seconds. \label{tb:training_time-MACE}}
    \centering
    \begin{tabular}{crr}
    \toprule
                & Baseline  & PIWSL  \\
    \cmidrule(lr){1-1} \cmidrule(lr){2-3}
    MACE-OFF & 18.5   & 31.7       \\
    \bottomrule
    \end{tabular}
\end{table}

\tabref{tb:training_time} provides training times measured for experiments with and without PIWSL. The training time is measured for a single training epoch and is averaged over five epochs in total. The experiments were performed using 1000 training configurations from the ANI-1x data set. We used a mini-batch of six. The table indicates that PIWSL increases the training time by a factor of two to three compared to the baseline (due to the additional gradient calculations). This is primarily because our PITC and PISC losses effectively double or triple the number of data labels, resulting in a proportional increase in training time due to the expanded set of training labels. This issue can be mitigated by employing the 2pt-PISC loss introduced in \secref{sec:PISC-variance}, which eliminates the need to estimate the third conformation. \tabref{tb:training_time-MACE} presents the average training times per epoch for MACE-OFF over ten epochs, as discussed in \autoref{sec:md17cc-finetune}. These results indicate that even the MACE-OFF (large) model requires less than twice the training time while achieving approximately a 40\% reduction in error. We emphasize that the PIWSL approach only alters training time; the inference time is unaffected.

\subsection{Different Configurations for the Physics-Informed Spatial-Consistency Loss \label{sec:other-consistency}}

\subsubsection{Triangle-Based}

\begin{table*}[h]
    \caption{\textbf{Results for different configurations of the PISC loss.} The presented numerical values are the root mean square errors (RMSEs) for the ANI-1x data set~\citep{smith2020ani}. Energy (in kcal/mol) and force (in kcal/mol/\AA) errors are obtained by averaging over three independent runs. All models are trained using 1000 configurations. The case 1, 2, and 3 correspond to \eqref{eq:spatial-consis-loss}, \eqref{eq:SC2} and \eqref{eq:SC3}, respectively. \label{tb:Ani1x_consis_results}}
    \centering
    \begin{tabular}{lllrrrr}
    \toprule
    Model                     &   & Baseline  & PISC (Case 1) & PISC (Case 2)     & PISC (Case 3)     \\
    \cmidrule(lr){1-2} \cmidrule(lr){3-6}
    \multirow[c]{2}{*}{PaiNN} & E & 60.11     & {\bf 45.24}   & 46.32             & 57.29             \\
                                       & F & 13.10     & {\bf 12.33}   & 12.42             & 13.28             \\
    \bottomrule
    \end{tabular}
\end{table*}

In \secref{sec:conservative-force}, we consider the following form of the PISC loss
\begin{equation}
    \label{eq:spatial-consis-loss-again}
    \begin{split}
        L_\mathrm{PISC}\left(\mathcal{S};\boldsymbol{\theta}\right) & = \ell\left( E\left(\mathcal{S}_{\dr}; \boldsymbol{\theta}\right), E\left(\mathcal{S}_{\dr^\prime}; \boldsymbol{\theta}\right) - \sum_{i=1}^{N_\mathrm{at}} \langle \dr_i^{\prime\prime}, \mathbf{F}_i\left(\mathcal{S}_{\dr^\prime}; \boldsymbol{\theta}\right) \rangle \right),
    \end{split}
\end{equation}
where $\dr, \dr', \dr''$ are related as $\dr' + \dr'' = \dr$. In this section, as a variant of \eqref{eq:spatial-consis-loss}, we also consider the following three PISC losses
\begin{align}
  L_\mathrm{PISC,Case\,2}\left(\mathcal{S};\boldsymbol{\theta}\right) & = \ell\left( E\left(\mathcal{S}; \boldsymbol{\theta}\right) - \sum_{i=1}^{N_\mathrm{at}} \langle \dr_i, \mathbf{F}_i\left(\mathcal{S}; \boldsymbol{\theta}\right) \rangle, E\left(\mathcal{S}_{\dr^\prime}; \boldsymbol{\theta}\right) - \sum_{i=1}^{N_\mathrm{at}} \langle \dr_i^{\prime\prime}, \mathbf{F}_i\left(\mathcal{S}_{\dr^\prime}; \boldsymbol{\theta}\right) \rangle \right),
  \label{eq:SC2}
  \\
  L_\mathrm{PISC,Case\,3}\left(\mathcal{S};\boldsymbol{\theta}\right) & = \ell\left( E\left(\mathcal{S}_{\dr'}; \boldsymbol{\theta}\right), E\left(\mathcal{S}_{\dr}; \boldsymbol{\theta}\right) - \sum_{i=1}^{N_\mathrm{at}} \langle -\dr_i^{\prime\prime}, \mathbf{F}_i\left(\mathcal{S}_{\dr}; \boldsymbol{\theta}\right) \rangle \right),
  \label{eq:SC3}     
\end{align}
where the point at $\mathbf{r} + \dr$ is the point where PIRC loss is imposed (see \eqref{eq:taylor-wsl-loss}). The results are provided in \tabref{tb:Ani1x_consis_results} and indicate that \eqref{eq:spatial-consis-loss} (Case 1) shows a better performance than the other cases for both the potential energy and the force predictions. 
In this study, we used the ANI-1x data set with 1000 training samples different from the one used to train the model used in the main body to avoid overfitting on the test data set. For the coefficient of the PITC and PISC losses, we used $0.1$ and $0.001$ with $\epsilon_{\rm max} = 0.01$.

\subsubsection{Further Variations\label{sec:PISC-variance}}

\paragraph{Two-Point Spacial Consistency.}
The flexibility of the PISC loss allows us to explore additional forms of spatial consistency. For example, instead of using a triangular configuration, we can impose spatial consistency between two points at $\mathbf{r}$ and $\mathbf{r} + \dr$, leading to the following expression: 
\begin{equation}
    \label{eq:spatial-2pt-consis-loss-with-label}
    \begin{split}
        L_\mathrm{PISC,2pt}\left(\mathcal{S};\boldsymbol{\theta}\right) & = \ell\left( E\left(\mathcal{S}; \boldsymbol{\theta}\right), E\left(\mathcal{S}_{\dr}; \boldsymbol{\theta}\right) - \sum_{i=1}^{N_\mathrm{at}} \langle -\dr_i, \mathbf{F}_i\left(\mathcal{S}_{\dr}; \boldsymbol{\theta}\right) \rangle \right). 
    \end{split}
\end{equation}
While not thoroughly investigated, we empirically observed that this loss delivers competitive performance when applied with the same coefficient value as the PITC loss.

\paragraph{Two-Point Spacial Consistency with Label.}
Since the training sample includes a label, we can utilize this label information instead of predicting the potential energy of the original conformation to enforce two-point spatial consistency. 
\begin{equation}
    \label{eq:spatial-2pt-consis-loss}
    \begin{split}
        L_\mathrm{PISC,2ptwl}\left(\mathcal{S};\boldsymbol{\theta}\right) & = \ell\left(E_{\mathcal{S}}^{\rm ref} , E\left(\mathcal{S}_{\dr}; \boldsymbol{\theta}\right) - \sum_{i=1}^{N_\mathrm{at}} \langle -\dr_i, \mathbf{F}_i\left(\mathcal{S}_{\dr}; \boldsymbol{\theta}\right) \rangle \right). 
    \end{split}
\end{equation}
We observed that this loss function performs well when applying the MACE model to datasets without force labels, as discussed in \autoref{sec:md17cc-finetune}.

\paragraph{Second-order Term Consideration for PITC Loss.}
In \eqref{eq:taylor-wsl}, we considered only the first-order Taylor expansion. Here, we introduce a straightforward approach to approximately account for the second-order term. Using the energy and forces at ${\bf r} + \dr$, the second-order derivative of the potential energy can be approximated using the explicit finite difference method as:
\begin{equation}
   \dr^{\mu} \dr^{\nu} \partial_{\mu} \partial_{\nu} E = \dr^{\mu} \dr^{\nu} \partial_{\mu} F_{\nu} \simeq \langle \dr, \mathbf{F}\left(\mathcal{S}_{\dr}; \boldsymbol{\theta}\right) - \mathbf{F}\left(\mathcal{S}; \boldsymbol{\theta}\right) \rangle, 
\end{equation}
Then, \eqref{eq:taylor-wsl} with the above second-order term can be written as: 
\begin{align}
    \label{eq:taylor-2nd-wsl}
        E\left(\mathcal{S}_{\dr}; \boldsymbol{\theta}\right) & \approx  E\left(\mathcal{S}; \boldsymbol{\theta}\right) 
        \nonumber
        \\
        &- \sum_{i=1}^{N_\mathrm{at}} \left[ \langle \dr_i, \mathbf{F}_i \left(\mathcal{S}; \boldsymbol{\theta}\right) \rangle + k_{\rm 2nd} \langle \dr_i, \mathbf{F}_i\left(\mathcal{S}_{\dr}; \boldsymbol{\theta}\right) - \mathbf{F}_i\left(\mathcal{S}; \boldsymbol{\theta}\right) \rangle \right] + \mathcal{O}\left(\|\dr\|^3\right),
        \nonumber
        \\
        & = E\left(\mathcal{S}; \boldsymbol{\theta}\right) 
        - \sum_{i=1}^{N_\mathrm{at}} \langle \dr_i, (1 - k_{\rm 2nd}) \mathbf{F}_i \left(\mathcal{S}; \boldsymbol{\theta}\right) + k_{\rm 2nd} \mathbf{F}_i\left(\mathcal{S}_{\dr}; \boldsymbol{\theta}\right) \rangle + \mathcal{O}\left(\|\dr\|^3\right),
\end{align}
where $k_{\rm 2nd}$ is a numeric parameter that controls the contribution of the second-order term. Note that setting $k_{\rm 2nd} = 0.5$ recovers the exact second-order expression, while $k_{\rm 2nd} = 0$ reverts to the first-order approximation. 

Using \eqref{eq:taylor-2nd-wsl}, the expression of the second-order PITC loss is given as: 
\begin{equation}
    \label{eq:taylor-2nd-wsl-loss}
    \begin{split}
        L_\mathrm{PITC}\left(\mathcal{S};\boldsymbol{\theta}\right) & =  \ell\Big( E\left(\mathcal{S}_{\dr}; \boldsymbol{\theta}\right), E\left(\mathcal{S}; \boldsymbol{\theta}\right) - \sum_{i=1}^{N_\mathrm{at}} \langle \dr_i, (1 - k_{\rm 2nd}) \mathbf{F}_i \left(\mathcal{S}; \boldsymbol{\theta}\right) + k_{\rm 2nd} \mathbf{F}_i\left(\mathcal{S}_{\dr}; \boldsymbol{\theta}\right) \rangle \Big),
    \end{split}
\end{equation}

Note that the second-order term becomes significant under the following conditions: (1) the accuracy of the MLIP model surpasses the contribution of the first-order term, and (2) the original conformation is near the equilibrium state. The latter can be understood as follows. Considering a spring model as a two-body interaction, the potential energy can be expressed as:
\begin{equation}
    V({\bf r}) = \frac{k_{\rm bond}}{2} ({\bf r} - {\bf r}_0)^2, 
\end{equation}
where $k_{\rm bond}$ is the constant characterizing the strength of the two-body interaction, and ${\rm r_0}$ denotes the equilibrium bond length. Introducing a small perturbation $\dr$ in the bond length, the potential energy becomes:
\begin{equation}
    V({\bf r} + \dr) = \frac{k_{\rm bond}}{2} ({\bf r} - {\bf r}_0 + \dr)^2 = V({\bf r}) + 2 k_{\rm bond} \langle {\bf r} - {\bf r}_0, \dr \rangle + \frac{k_{\rm bond}}{2} \dr^2.
\end{equation}
At the equilibrium state (${\bf r} = {\bf r}_0$), the above equation demonstrates that the second-order term becomes dominant. Consequently, the second-order accuracy of the PITC loss function becomes crucial in such scenarios.

\paragraph{PITS for Curl of Forces} 

Another potential direction is enforcing a reduction in the curl of the forces. This can be achieved by leveraging Stokes' theorem: $\int_{\Sigma} \nabla \times \mathbf{F} \cdot d \mathbf{S} = \oint_{\partial \Sigma} \mathbf{F} \cdot d\mathbf{l} = 0$ where $\Sigma$ represents a specific surface regime, and $\partial \Sigma$ denotes its boundary. Similar to the PISC loss, the right-hand side of this equation can be effectively described by considering a triangular configuration, where the midpoints of the three sides correspond to $\mathbf{r}, \mathbf{r} + \dr, \mathbf{r} + \dr^{\prime}$. 

\color{black}
\subsection{Detailed Setups for Qualitative Analysis}
\subsubsection{C--H Potential Energy Profile of Aspirin \label{sec:vis-setup}}

\begin{table}[h]
    \caption{\textbf{Performance of PaiNN employed in \figref{fig:enter-label} (c, d).} All the models other than the reference model ($N_{\rm train} = 1000$) use the OC20's hyper-parameters. For the reference model, we tuned the hyper-paramters of PaiNN model following the original paper \citep{schutt2021equivariant}.}
    \label{tb:PaiNN-Aspirin-CH}
    \begin{center}
    \begin{tabular}{llccccc}
    \toprule
                              &   & \multicolumn{2}{c}{$N_\mathrm{train} = 100$}    & \multicolumn{2}{c}{$N_\mathrm{train} = 200$}  & $N_\mathrm{train} = 1000$     \\    
                              &   & Baseline    & PIWSL                             & Baseline  & PIWSL                             & Baseline                      \\
    \cmidrule(lr){1-2} \cmidrule(lr){3-4} \cmidrule(lr){5-6} \cmidrule(lr){7-7}
    \multirow[c]{2}{*}{PaiNN} & E & 6.55        & {\bf 5.64}                        & 5.11      & {\bf 4.48}                        & 0.68                          \\
                              & F & 7.38        & 7.36                              & 3.95      & 3.97                              & 1.44                          \\
    \bottomrule
    \end{tabular}
    \end{center}
\end{table}

This section describes the detailed setup and procedure for \secref{sec:vis-PIWSL}. First, we trained PaiNN with and without PIWSL losses using the aspirin data from rMD17 with training set sizes of 100 and 200. For PIWSL, we used $(C_{\rm PITC}, C_{\rm PISC}, \epsilon_{\rm max}) = (1.2, 0.01, 0.015)$. The other experimental setups are the same as for rMD17 experiments presented in \secref{sec:setup}. We used the PaiNN model with gradient-based forces to obtain the reference model and tuned the model hyper-parameter with Optuna \citep{optuna_2019}. The obtained models' performance is provided in \tabref{tb:PaiNN-Aspirin-CH}. Then, we prepared the aspirin molecule structures, including the corresponding atomic coordinates and atomic types. For these structures, we perturbed one of the C-H bonds with a bond length from 0.8 \AA{} to 1.8 \AA{}. We prepared 100 structures and estimated the corresponding potential energy with the pre-trained models. The aspirin data is provided in our publicly available source code.

\subsubsection{MD Simulation Stability Analysis \label{sec:md-setup}}

\paragraph*{NVE-Ensembles} This section describes the detailed setup and the procedure for our analysis of MD simulations in \secref{sec:vis-PIWSL}. Because our implementation builds upon the source code provided by \citet{fu2023forces}, we used their scripts for performing MD simulations. However, we added a minor modification to enable MD simulations in the microcanonical (NVE) statistical ensemble, i.e., the particle position and velocity are updated with velocity Verlet algorithm\citep{verlet1967computer}~\footnote{Note that the total energy conservation necessary for the microcanonical statistical ensemble is in general not perfectly satisfied due to the numerical error, in particular, when the force is not calculated as the curl of the force.}. We set the initial temperature to 300~K and the integration time step to 0.5~fs for all simulations. As defined by \citet{fu2023forces}, the stability of an MD simulation for a target molecule is defined as the time $T$ during which the bond lengths satisfy the following condition
\begin{equation}
    \max_{(i,j) \in \mathcal{B} } |(||{\bf x}_i(T) - {\bf x}_j(T)||) - b_{i,j}| > \Delta\,,
    \label{eq:md-stability}
\end{equation}
where $\mathcal{B}$ denotes the set of all bonds, $\{i, j\}$ denote the pair of bonded atoms, and $b_{i, j}$ denotes the equilibrium bond length. Following \citet{fu2023forces}, we set $\Delta = 0.5 \AA$. This definition indicates when the molecule experiences significant structural changes during the MD simulation.

\begin{figure}[t]
    \centering
    \includegraphics[width=\textwidth]{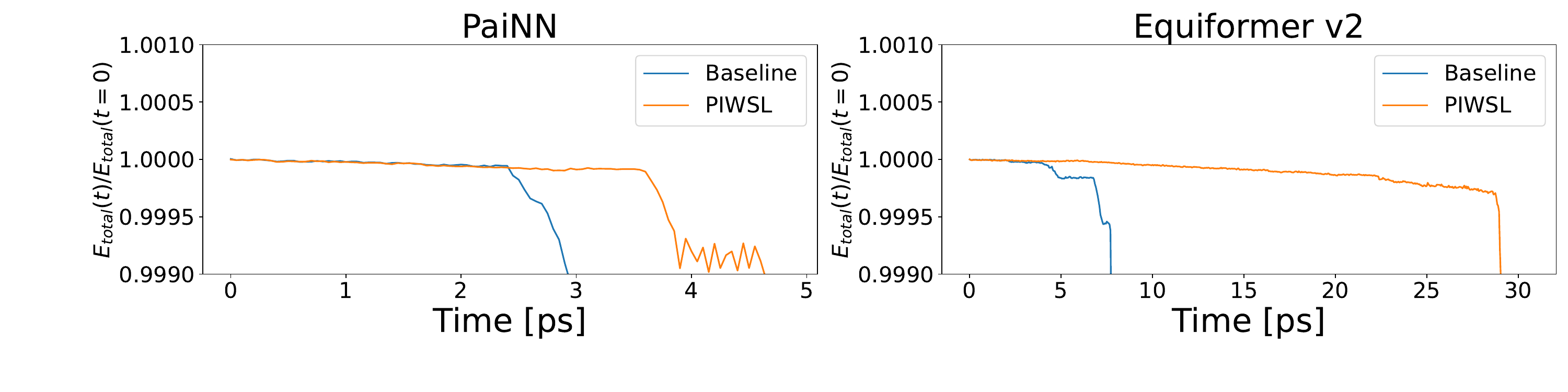}
    \caption{{\bf Analysis of the total energy conservation using MD simulations with MLIP models.} The amount of the change of the total energy during MD simulations is assessed for the baseline MLIP models and those trained with PIWSL. The total energy is measured at the initial and final time-step and the difference is normalized by the total energy at the initial time-step. All results are obtained for the aspirin molecule and MD simulations in the microcanonical (N V E) statistical ensemble.}
    \label{fig:E-conv-NVE}
\end{figure}

We trained PaiNN and Equiformer v2 models with and without PIWSL losses using the aspirin data from rMD17. We used training set sizes of 100 and 200 for PaiNN-GF and 1000 for PaiNN and Equiformer v2 with direct force. The corresponding stability values are presented in \tabref{tb:Stability-Aspirin-MD}. The hyperparameters for the PIWSL loss are provided in \tabref{tb:hyper-parameter-list-MD}. To train direct force model with 1000 samples, we used second-order PITC and PISC losses because of their good accuracy. The results for the stability of the PaiNN (direct and gradient-based force) and Equiformer v2 models are shown in \tabref{tb:RMSE-Aspirin-MD}. To select the hyperparameters in \tabref{tb:hyper-parameter-list-MD}, a series of MD simulations with a fixed random number seed for the initial atomic velocities was used. For our final stability results in \tabref{tb:Stability-Aspirin-MD}, a series of MD simulations with three different random seeds for velocities and the selected hyperparameters was used. 
\figref{fig:E-conv-NVE} presents the temporal evolution of the total energy during the MD simulations reported in \figref{fig:stability}, which shows that the total energy is conserved within 0.05\% deviation until the disruption. These results indicate that PIWSL slightly enhances the energy conservation capability of MLIP models in most cases, though the improvement is relatively modest due to the small deviations observed.

\begin{figure}[t]
    \centering
    \includegraphics[width=0.95\textwidth]{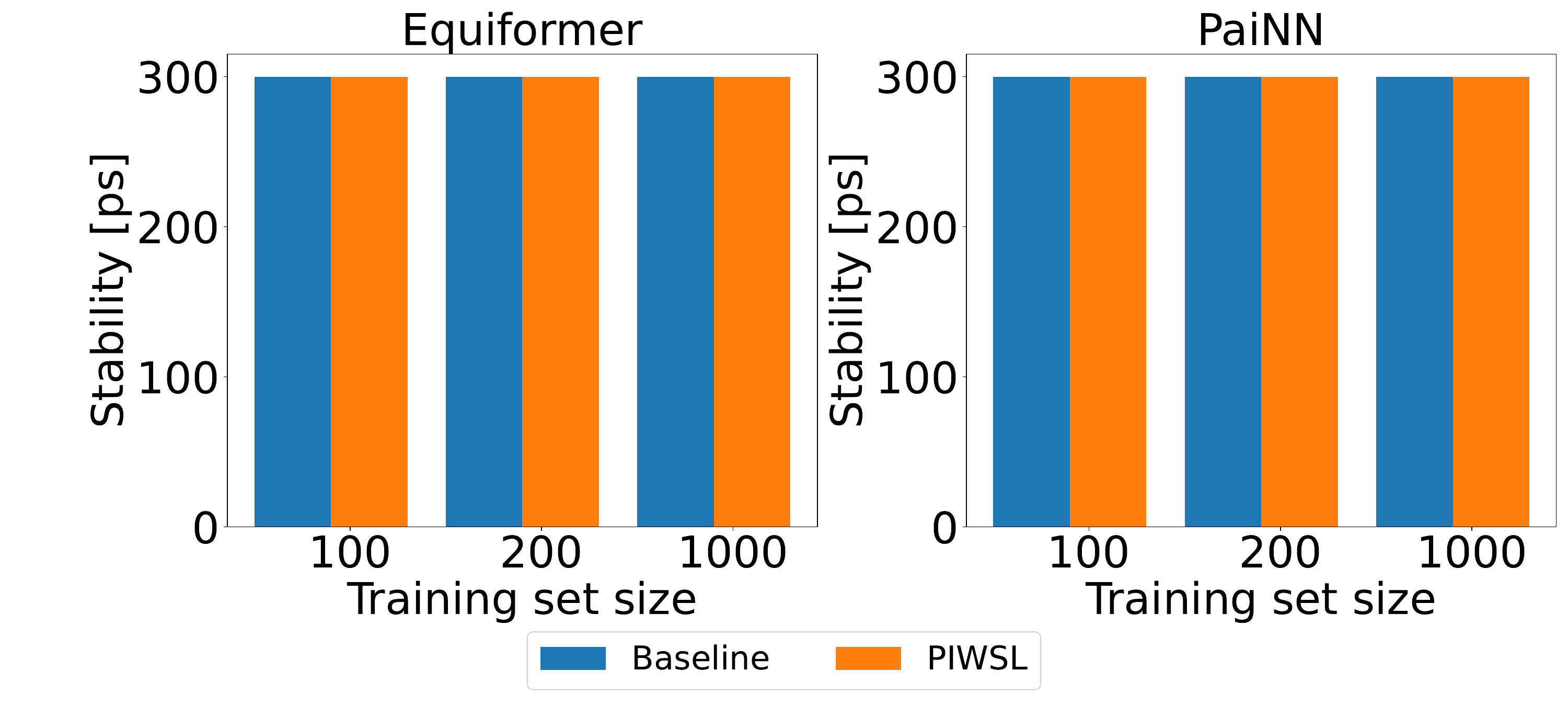}
    \caption{{\bf Stability analysis of the MLIP models during MD simulations.} Stability during MD
simulations is assessed for the baseline MLIP models and those trained with PIWSL. All results
are obtained for the aspirin molecule and MD simulations in the canonical (N V T) statistical
ensemble. We measure stability during MD simulations according to \citep{fu2023forces}.}
    \label{fig:NVT-stability}
\end{figure}

\paragraph*{NVT-Ensembles} To investigate the effect of thermostats, we also performed MD simulations in the canonical (NVT) statistical ensemble, where temperature is maintained constant. To keep the temperature constant, we used Nos\'e-Hoover thermostat~\citep{nose1984unified,hoover1985canonical}. The initial and target temperatures were both set to 300~K for all simulations. The integration time step was set to 0.5~fs and the characteristic parameter $\tau$ for the thermostat is set to 20~fs. The result, shown in \figref{fig:NVT-stability}, demonstrate that the thermostat stabilizes the simulations by mitigating the increase in kinetic energy. 

\begin{table}[h]
    \caption{\textbf{Stability of the models employed in the MD analysis.} The presented numerical values are the stability defined by \eqref{eq:md-stability} measured in ps. The results are obtained as an average over three different random seeds for the initial velocity of the atoms in the target aspirin molecule. "GF" denotes the gradient-based force prediction.}
    \label{tb:Stability-Aspirin-MD}
    \begin{center}
    \begin{tabular}{llccc}
    \toprule
    & & $N_\mathrm{train} = 100$    & $N_\mathrm{train} = 200$  & $N_\mathrm{train} = 1000$    \\    
    \cmidrule(lr){1-2} \cmidrule(lr){3-5}
    \multirow[c]{2}{*}{PaiNN} 
    & Baseline & -- & -- & 2.68 $\pm$ 0.13 \\
    & PIWSL    & -- & -- & {\bf 4.65 $\pm$ 0.72} \\
    \midrule                          
    \multirow[c]{2}{*}{Equiformer} 
    & Baseline & -- & -- & 14.53 $\pm$ 8.61 \\
    & PIWSL & -- & -- & {\bf 24.65 $\pm$ 8.24} \\
    \midrule
    \multirow[c]{2}{*}{PaiNN-GF} 
    & Baseline & {\bf 3.25 $\pm$ 3.98} & {\bf 220.5 $\pm$ 137.7} & -- \\
    & PIWSL & {\bf 15.07 $\pm$ 10.09} & {\bf 267.7 $\pm$ 56.0} & -- \\
    \bottomrule
    \end{tabular}
    \end{center}
\end{table}

\begin{table}[h]
    \caption{\textbf{Hyper-parameters for the PIWSL loss used in the MD analysis.} We used the following hyper-parameter for MD simulation analysis: ($C_{\rm PITC}, C_{\rm PISC}, \epsilon_\mathrm{max})$= Case $\alpha$: (0.01, 0.001, 0.025), Case $\beta$: (1.2, 0.01, 0.01), Case $\gamma$: (1.2, 0.01, 0.025), Case $\delta$: (1.2, 0.01, 0.015), Case $\epsilon$: (0.1, 0.01, 0.01), and Case $\zeta$: (1.0, 0., 0.01). "GF" denotes the gradient-based force prediction. \label{tb:hyper-parameter-list-MD}}
    \centering
    \begin{tabular}{lrccc}
    \toprule
    Dataset         & Size      & Equiformer v2 & PaiNN & PaiNN-GF \\
    \cmidrule(lr){1-2} \cmidrule(lr){3-5}
    rMD17           & 100      & $\alpha$  & $\delta$ & $\epsilon$\\
    (Aspirin)       & 200      & $\beta$   & $\beta$ & $\zeta$ \\
                    & 1000      & $\epsilon$  & $\gamma$ & -- \\
    \bottomrule
    \end{tabular}
\end{table}

\begin{table}[h]
    \caption{\textbf{Energy and force errors for the models employed in the MD analysis.} The presented numerical values are the root-mean-square errors (RMSEs) of energy (E) and force (F). Energy RMSE is given in kcal/mol, while force RMSE is in kcal/mol/\AA. "GF" denotes the gradient-based force prediction. \label{tb:RMSE-Aspirin-MD}}
    \begin{center}
    \begin{tabular}{llcccccc}
    \toprule
                              &   & \multicolumn{2}{c}{$N_\mathrm{train} = 100$}    & \multicolumn{2}{c}{$N_\mathrm{train} = 200$}  & \multicolumn{2}{c}{$N_\mathrm{train} = 1000$}    \\    
                              &   & Baseline    & PIWSL                             & Baseline  & PIWSL                             & Baseline  & PIWSL         \\
    \cmidrule(lr){1-2} \cmidrule(lr){3-4} \cmidrule(lr){5-6} \cmidrule(lr){7-8}
    \multirow[c]{2}{*}{PaiNN-DF} & E & 6.55 & 5.64    & 5.11 & 4.48   & 2.30 & 0.99 \\
                              & F & 7.38 & 7.36    & 3.95 & 3.97   & 1.63 & 1.61  \\
    \cmidrule(lr){1-2} \cmidrule(lr){3-4} \cmidrule(lr){5-6} \cmidrule(lr){7-8}            
    \multirow[c]{2}{*}{Equiformer v2} & E & 4.79 & 4.64  & 4.92 & 4.82   & 1.39 & 1.19 \\
                                   & F & 4.86 & 4.90  & 2.50 & 2.42   & 0.76 & 0.75 \\ 
    \midrule                               
    \multirow[c]{2}{*}{PaiNN-GF} & E & 6.05 & 6.03    & 6.01 & 6.02   & -- & -- \\
                              & F & 6.41 & 6.33    & 3.50 & 3.53   & -- & --  \\
    \bottomrule
    \end{tabular}
    \end{center}
\end{table}

\subsection{Training setup for MD17-CCSD(T) Experiments\label{sec:md17cc-setup}}

In this section, we provide the training setup of finetuning of MACE-OFF discussed in \secref{sec:md17cc-finetune}. 
The considered model is MACE-OFF (large) \citep{Kovacs2023} which is the MACE model pretrained on SPICE \citep{eastman2023spice}, QMugs \citep{isert2022qmugs}, and liquid water \citep{schran2021machine} datasets. 
The foundation model is finetuned on the aspirin molecule data in CCSD dataset \citep{chmiela2018towards} whose potential energy is obtained at quantum-chemical CCSD level of accuracy. The data includes 1500 samples which are splitted into 950/50/500 as train/validation/test datasets. To emulate general coupling-cluster method dataset, we only use potential energy label for training. We utilize the set up provided in the official repository\footnote{\url{https://mace-docs.readthedocs.io/en/latest/guide/finetuning.html}} with modifying learning rate from $10^{-2}$ to $10^{-3}$, to improve the performance. For the model trained scratch, we use the same architecture with setting energy weight to 40 and learning rate to $10^{-3}$ to improve the performance. For PIWSL, we used the second-order PITC and 2pt-with-label PISC losses \autoref{eq:spatial-2pt-consis-loss-with-label} with the following parameters: ($C_{\rm PITC}, C_{\rm PISC}, \epsilon) = (0.55, 4.5, 0.01)$. 

\subsection{Metric Dependence of PITC \label{sec:neg-small-term}}

\begin{table*}[h]
    \caption{\textbf{Metric dependence of PITC.} The presented numerical values are the root mean square errors (RMSEs) for the ANI-1x data set~\citep{smith2020ani}. Energy (in kcal/mol) and force (in kcal/mol/\AA) errors are obtained by averaging over three independent runs. All models are trained using 1000 configurations. MAE refers to the mean absolute error, and MSE denotes the mean square error. \label{tb:Ani1x_ReLU_results}}
    \centering
    \begin{tabular}{llrrrrr}
    \toprule
    Model                     &   & Baseline  & PITC MAE Loss     & PITC MSE Loss     & PITC ReLU Loss    \\
    \cmidrule(lr){1-2} \cmidrule(lr){3-6}
    \multirow[c]{2}{*}{PaiNN} & E & 60.11     & 58.84             & {\bf 47.09}       & 60.47             \\
                              & F & 13.10     & 13.18             & {\bf 12.19}       & 13.06             \\
    \bottomrule
    \end{tabular}
\end{table*}

\Tabref{tb:Ani1x_ReLU_results} provides the result of the metric dependence of PIWSL. For simplicity, we only consider the PITC loss (the coefficient of the PITC and PISC losses are set as $0.1$ and $0$). For the ReLU metric, we consider
\begin{equation}
    \begin{split}
        L_\mathrm{\rm ReLU}\left(\mathcal{S};\boldsymbol{\theta}\right) & = \mathrm{ReLU}\left( \left| E\left(\mathcal{S}; \boldsymbol{\theta}\right) - \sum_{i=1}^{N_\mathrm{at}} \langle \dr_i, \mathbf{F}_i\left(\mathcal{S}; \boldsymbol{\theta}\right) \rangle -  E\left(\mathcal{S}_{\dr}; \boldsymbol{\theta}\right) \right| - E\left(\mathcal{S}_{\dr}; \boldsymbol{\theta}\right) ||\dr||^2 \right).
    \end{split}
\end{equation}
This metric is zero when the difference between the two terms is less than the second-order term in $\dr$. The results indicate that taking the second-order term into account does not improve the performance (see PITC MAE Loss and PITC ReLU Loss results), and the MSE loss function shows the best performance. In this study, we used the ANI-1x data set and the 1000 training samples. These samples differ from the one used to train the model in the main text to avoid overfitting the test data set.

\subsection{Perturbation Magnitude Dependence of PITC}

\begin{table*}[h]
    \caption{\textbf{Perturbation magnitude dependence of PITC.} The presented numerical values are the root mean square errors (RMSEs) for the ANI-1x data set~\citep{smith2020ani}. Energy (in kcal/mol) and force (in kcal/mol/\AA) errors are obtained by averaging over three independent runs. All models are trained using 1000 configurations. \label{tb:Ani1x_length_results}} 
    \centering
    \begin{tabular}{llrrrr}
    \toprule
    Model                         &   & Baseline  & $\epsilon_{\rm max} = 0.0005$ & $\epsilon_{\rm max} = 0.005$  & $\epsilon_{\rm max} = 0.05$   \\
    \cmidrule(lr){1-2} \cmidrule(lr){3-6}
    \multirow[c]{2}{*}{PaiNN}     & E & 60.11     & 60.43                         & {\bf 47.09}                   & 109.17                        \\
               & F & 13.10     & 12.75                         & 12.19                         & {\bf 11.70}                   \\
    \bottomrule
    \end{tabular}
\end{table*}

In this section, we provide the result of the perturbation magnitude dependence of PIWSL, i.e., $\|\dr\| = \epsilon$. For simplicity, we only consider the PITC loss (the coefficient of the PITC and PISC losses are set as $0.1$ and $0.0$). The results are provided in \tabref{tb:Ani1x_length_results} and demonstrate that the longer perturbation vector length is fruitful for force predictions. However, values that are too large are harmful to predicting potential energy. In this study, we used the ANI-1x data set and the 1000 training samples. These samples differ from the one used to train the model in the main text to avoid overfitting the test data set.

\subsection{Dependence of PITC on the Number of Perturbed Atoms}

\begin{table*}[h]
    \caption{\textbf{Dependence of PITC on the number of perturbed atoms.} The presented numerical values are the root mean square errors (RMSEs) for the ANI-1x data set~\citep{smith2020ani}. Energy (in kcal/mol) and force (in kcal/mol/\AA) errors are obtained by averaging over three independent runs. All models are trained using 1000 configurations. \label{tb:Ani1x_anumber_results}}
    \centering
    \begin{tabular}{llrrrrrrr}
    \toprule
    Model                     &   & Baseline  & 10\%  & 20\%  & 50\%  & 75\%          & 90\%          & 100 \%    \\
    \cmidrule(lr){1-2} \cmidrule(lr){3-9}
    \multirow[c]{2}{*}{PaiNN} & E & 60.11     & 46.68 & 52.37 & 54.51 & 46.94         & {\bf 45.92}   & 46.32     \\
                              & F & 13.10     & 13.03 & 12.62 & 12.16 & {\bf 12.14}   & 12.24         & 12.42     \\
    \bottomrule
    \end{tabular}
\end{table*}

This section provides the result of the perturbed atom number dependence of PIWSL. For simplicity, we only consider the PITC loss (the coefficient of the PIRC and PISC losses are set as $0.1$ and $0$). In this study, we randomly selected atoms in a training sample following the ratio of $0~\%, 10~\%, 20~\%, 50~\%, 75~\%, 90~\%, 100~\%$. The results are provided in \tabref{tb:Ani1x_anumber_results}, which indicates that around 75\% to 100\% ratio cases result in the best performance for the force and the potential energy prediction. However, the number dependence is rather complicated. Therefore, in the main text, we perturbed all the atoms (100 \%) as a conservative choice. In this study, we used the ANI-1x data set and the 1000 training samples. These samples differ from the one used to train the model in the main text to avoid overfitting the test data set.

\subsection{Dependence on the Number of Training Iterations \label{sec:iter-study}}

\begin{table}[h]
    \caption{\textbf{Dependence on the number of training iterations.} The presented numerical values are the root mean square errors (RMSEs) for the ANI-1x data set~\citep{smith2020ani}. Energy (in kcal/mol) and force (in kcal/mol/\AA) errors are obtained by averaging over three independent runs. All models are trained using 1000 configurations. \label{tb:Ani1x_iter_results}}
    \centering
    \begin{tabular}{lllrrr}
    \toprule
    Model                       & Iteration Number  &   & Baseline          & PIWSL                     \\
    \cmidrule(lr){1-1} \cmidrule(lr){2-3} \cmidrule(lr){4-5}
    \multirow[c]{4}{*}{PaiNN}   & 40,000            & E & 56.62 $\pm$ 2.80  & {\bf 24.53 $\pm$ 0.16}    \\
                                &                   & F & 12.96 $\pm$ 0.18  & {\bf 11.43 $\pm$ 0.05}    \\
    \cmidrule(lr){2-3} \cmidrule(lr){4-5}
                                & 80,000            & E & 59.92 $\pm$ 1.47  & {\bf 23.78 $\pm$ 0.16}    \\
                                &                   & F & 13.10 $\pm$ 0.19  & {\bf 11.50 $\pm$ 0.04}    \\
    \bottomrule
    \end{tabular}
\end{table}

To show the effectiveness of our approach even in the case of longer training, we provide the result of the dependence of PIWSL on the number of training iterations. In this study, we performed training twice as long as in the main text, that is, 80,000 iterations for ANI-1x with 1000 training samples. The results are provided in \tabref{tb:Ani1x_iter_results} and indicate that our approach performs better in the longer training case. On the other hand, the training without PIWSL shows an overfitting to the validation data set, reducing its performance compared to the shorter training case. In this study, we used the ANI-1x data set and the 1000 training samples. These samples differ from the one used to train the model in the main text to avoid overfitting the test data set. The coefficients of the PITC and PISC losses are $1.2$ and $0.01$, respectively.

\subsection{Additional Experiments with Gradient-Based Forces \label{sec:GF-experiments}}

\begin{table}[h]
    \caption{\textbf{Results of PIWSL with gradient-based force predictions.} The presented numerical values are the root mean square errors (RMSEs) for the ANI-1x data set~\citep{smith2020ani}. Energy (in kcal/mol) and force (in kcal/mol/\AA) errors are obtained by averaging over three independent runs. All models are trained using 1000 configurations. \label{tb:Ani1x_GF_results}}
    \centering
    \begin{tabular}{llrrrrr}
    \toprule
    Model & & Baseline (GF) & PIWSL (GF) & WL (GF) \\
    \cmidrule(lr){1-2} \cmidrule(lr){3-5}
    \multirow[c]{2}{*}{PaiNN}
    & E & 23.57 $\pm$ 0.62 & {\bf 18.62 $\pm$ 0.09} & 22.61 $\pm$ 0.50  \\
    & F & 11.32 $\pm$ 0.08 & {\bf 10.94 $\pm$ 0.01} & 11.72 $\pm$ 0.06 \\
    \cmidrule(lr){1-2} \cmidrule(lr){3-5}
    \multirow[c]{2}{*}{Equiformer}
    & E & 29.07 $\pm$ 2.32 & {\bf 19.53 $\pm$ 0.32} & 21.07 $\pm$ 0.86 \\
    & F & {\bf 11.90 $\pm$ 0.13} & {\bf 11.99 $\pm$ 0.03} & {\bf 11.90 $\pm$ 0.20} \\
    \bottomrule
    \end{tabular}
\end{table}

In this section, we provide the result of the training with the gradient-based force predictions. The results are provided in \tabref{tb:Ani1x_GF_results} and demonstrate that our PIWSL loss enables a better force prediction, even in the case of gradient-based force predictions. These results also indicate that our PIWSL method can improve the ML model performance in the case of MLIPs commonly applied in computational chemistry and materials science. We consider that this is partly due to the effectiveness of the weak label at $\mathbf{r} + \dr$ as indicated by the WL results, which show an improvement of the performance different from the case with the direct force branch (see also \secref{sec:analysis}). We hypothesize that the further improvement results from the additional gradient calculation as indicated in \eqref{eq:dif-taylor-wsl-loss} and \eqref{eq:dif-sp-wsl-loss}. This observation also indicates that our PIWSL method can potentially improve other generic property prediction tasks by calculating their first derivatives in terms of the atomic coordinate and utilizing the proposed loss functions. In this study, the coefficient of the PITC and PISC losses are set as $0.1$ and $0.01$ with $\epsilon_{\rm max} = 0.01$. The weak label loss coefficient is set as $0.1$.

\subsection{Reducing Curl of Forces for Models with the Force Branch \label{sec:force-rot-experiment}}

\begin{table}[h]
    \caption{\textbf{Curl of forces for models with the force branch.} The presented numerical values are the absolute values of the total curl of the force evaluated for the ANI-1x data set~\citep{smith2020ani}. Energy (in kcal/mol) and force (in kcal/mol/\AA) errors are obtained by averaging over three independent runs. All models are trained using 1000 configurations. \label{tb:Ani1x_rotF_results}}
    \centering
    \begin{tabular}{lrr}
    \toprule
    Model       & Baseline          & PITC                      \\
    \cmidrule(lr){1-1} \cmidrule(lr){2-3} 
    PaiNN       & 45.18 $\pm$ 4.07  & {\bf 39.06 $\pm$ 0.58}    \\
    Equiformer  & 29.62 $\pm$ 0.28  & {\bf 23.42 $\pm$ 0.09}    \\
    \bottomrule
    \end{tabular}
\end{table}

In this section, we study the effect of our loss functions on the curl of forces in the case of the model with the force branch. The results are provided in \tabref{tb:Ani1x_rotF_results}, which shows that our PITC loss reduces the curl of the predicted forces, allowing potentially better energy conservation during MD simulations. In this study, we used the ANI-1x data set and the 1000 training samples. These samples differ from the one used to train the model in the main text to avoid overfitting the test data set. The hyper-parameters of the PITC and PISC losses are $(C_{\rm PITC}, C_{\rm PISC}, \epsilon_{\rm max}) = (1.2, 0.01, 0.025)$.
It is theoretically possible to define a loss function aimed at directly minimizing the absolute value of the curl of forces. However, this approach necessitates calculating the Hessian matrix, which requires a substantial memory cost given the limitations of current computational resources. Developing a method to train with such a loss function while mitigating memory requirements is a promising direction for future research. 

\section{Further Analyses of PIWSL \label{sec:analysis}}

The following provides further analyses of our approach. We provide the results for Equiformer v2 and PaiNN since these models employ equivariant features and demonstrate a high accuracy on the ANI-1x data set when trained using 1000 configurations.

\subsection{Training MLIPs without Reference Forces \label{sec:no-force-label}}

In the following, we explore scenarios where only potential energy labels are available. This situation commonly arises when calculating energy labels with chemically accurate approaches, such as CCSD(T)/CBS~\citep{Hobza2002, Feller2006}, for which force calculation is infeasible. To consider practical applications, we examine two cases: (1) predicting force by a force branch (FB) and (2) predicting force as a gradient of the potential energy (GF). The former enables fast force prediction and is popular in the machine learning community, while the latter requires additional gradient calculation but yields curl-free force predictions. It is popular in computational chemistry as it ensures the conservation of the total energy during MD simulations. The results are provided in \tabref{tb:Ani1x_noFL_results}; training without reference forces is achieved by setting the relative force contribution to zero in \eqref{eq:MLIP-loss}. The PIWSL method consistently performs better than the baseline for the FB and GF cases. However, a more significant improvement in the force prediction performance is observed in the GF case. We attribute this phenomenon to the inherent nature of PIWSL, which requires consistency between the potential energy and atomic forces, as discussed in \secref{sec:vis-PIWSL}. This result aligns with our expectations, confirming the capability of our PIWSL method to enable ML models to reduce the error in the predicted forces. Overall, PIWSL opens a new possibility for training MLIP models using highly accurate reference methods, such as CCSD(T)/CBS.

\begin{table}[t!]
    \captionof{table}{\textbf{Results for models trained on the ANI-1x data set without reference forces.} All models are trained using 1000 training samples. FB refers to the setting where the force branch estimates the force, and GF denotes the setting where the force is estimated by the gradient of the potential energy with respect to the atomic coordinates. \label{tb:Ani1x_noFL_results}}
    \begin{center}
    \resizebox{0.49\textwidth}{!}{
    \begin{tabular}{lccrr}
    \toprule
    Model                           & Case    &   & Baseline          & PIWSL                     \\
    \cmidrule(lr){1-3} \cmidrule(lr){4-5}
    \multirow[c]{4}{*}{PaiNN}       & FB      & E & 42.36 $\pm$ 0.30  & {\bf 25.42 $\pm$ 0.72}    \\
                                    &         & F & 24.25 $\pm$ 0.00  & {\bf 20.54 $\pm$ 0.08}    \\
                                    & GF      & E & 41.83 $\pm$ 1.81  & {\bf 29.71 $\pm$ 0.55}    \\
                                    &         & F & 83.36 $\pm$ 2.85  & {\bf 24.02 $\pm$ 0.95}    \\
    \cmidrule(lr){1-3} \cmidrule(lr){4-5}
    \multirow[c]{4}{*}{Equiformer}  & FB      & E & 43.14 $\pm$ 0.86  & {\bf 29.48 $\pm$ 0.51}    \\
                                    &         & F & 24.25 $\pm$ 0.00  & {\bf 21.99 $\pm$ 0.49}    \\
                                    & GF      & E & 42.55 $\pm$ 0.99  & {\bf 32.66 $\pm$ 1.11}    \\
                                    &         & F & 35.70 $\pm$ 0.78  & {\bf 21.83 $\pm$ 0.27}    \\
    \bottomrule
    \end{tabular}
    }
    \end{center}
\end{table}

\subsection{Comparing PITC with the Taylor-Expansion-Based Weak Label Loss.} 
We compare the PIWSL method with the Taylor-expansion-based weak label (WL) approach~\citep{cooper2020efficient}, whose loss function is presented in \eqref{eq:taylor-nong-loss}. For simplicity, we only consider the PITC loss in \eqref{eq:taylor-wsl-loss}. For a fair comparison, we consider the following two cases. First, we train with reference forces and energies (w. RF). Second, we train the methods without reference forces and use only the reference energies. For the training with reference forces, we set the numeric coefficient of the PITC loss to $1.0$; for the training without reference forces, the coefficient is set to $0.1$. Note that the WL loss without the reference force is calculated using the predicted force labels. The results are provided in \tabref{tb:Ani1x_cls_results}.

Our PITC loss demonstrates the best accuracy in all cases, with and without the reference forces. Interestingly, PaiNN failed to learn the potential energy with the WL loss and reference forces. We hypothesize this to be due to the imbalance of the training between the energies and forces. Specifically, the WL loss trains only the potential energy, resulting in an inconsistency between the energy and force branches, which share the same readout layer that experiences more frequent updates using the potential energy. This hypothesis is supported by the results for the training without reference forces, where the error in energy is reduced compared to the baseline. A further validation in a similar experiment in the case of GF is provided in \secref{sec:GF-experiments}. However, the proposed PITC loss still performs better here. In summary, the PITC loss enables MLIPs to learn energies and forces consistent with each other and does it better than the previously proposed WL method.

\begin{table}[t]
    \caption{\textbf{Comparison of PITC and the Taylor-expansion-based weak label loss.} WL (+FP) denotes the Taylor-expansion-based method using reference energies and either reference (w. RF) or predicted (w/o. RF) forces; see~\eqref{eq:taylor-nong-loss}. The listed values are the RMSE values for energies in kcal/mol and atomic forces in kcal/mol/\AA. All models are trained on the ANI-1x data set using 1000 configurations, with (w.) and without (w/o.) reference atomic forces (RF). \label{tb:Ani1x_cls_results}}
    \begin{center}
    \resizebox{0.7\columnwidth}{!}{
    \begin{tabular}{lllrrr}
    \toprule
    Model                           & Case      &   & Baseline          & PITC                  & WL (+FP) \\
    \cmidrule(lr){1-1} \cmidrule(lr){2-3} \cmidrule(lr){4-6}
    \multirow[c]{4}{*}{PaiNN}       & (w. RF)   & E & 56.62 $\pm$ 2.80  & {\bf 30.94$\pm$ 0.56} & 81.86$\pm$ 9.39   \\
                                    &           & F & 12.96$\pm$ 0.06   & {\bf 12.04$\pm$ 0.04} & 14.54$\pm$ 0.12   \\
    \cmidrule(lr){2-3} \cmidrule(lr){4-6}
                                    & (w/o. RF) & E & 42.36$\pm$ 0.30   & {\bf 25.42$\pm$ 0.72} & 41.77$\pm$ 4.82   \\
                                    &           & F & 24.25$\pm$ 0.00   & {\bf 20.54$\pm$ 0.08} & 24.68$\pm$ 0.54   \\
    \cmidrule(lr){1-1} \cmidrule(lr){2-3} \cmidrule(lr){4-6}
    \multirow[c]{4}{*}{Equiformer}  & (w. RF)   & E & 54.52$\pm$ 4.52   & {\bf 23.16$\pm$ 0.19} & 31.02$\pm$ 3.99   \\
                                    &           & F & 10.10$\pm$ 0.00   & {\bf 10.03$\pm$ 0.05} & 13.43$\pm$ 0.92   \\
    \cmidrule(lr){2-3} \cmidrule(lr){4-6}
                                    & (w/o. RF) & E & 43.14$\pm$ 0.86   & {\bf 29.48$\pm$ 0.51} & 88.59$\pm$ 11.36   \\
                                    &           & F & 24.25$\pm$ 0.00   & {\bf 21.99$\pm$ 0.49} & 293.41$\pm$ 26.96   \\
    \bottomrule
    \end{tabular}}
    \end{center}
\end{table}

\subsection{Ablating the Impact of PITC and PISC Losses.} We conduct an ablation experiment to analyze the impact of PITC and PISC losses. Results in \tabref{tb:Ani1x_ablation_results} indicate that the PITC loss predominantly improves the accuracy of resulting models, especially for PaiNN. Using just the PISC loss does not consistently improve accuracy but stabilizes training when combined with PITC. This combined approach notably benefits Equiformer v2. For Equiformer v2, we repeated the experiment five times to reduce the effect from an outlier on the PITC loss.

\begin{table}[t!]
    \captionof{table}{\textbf{Results for models with direct-force trained on the ANI-1x data set with ablated weakly supervised losses.} All models are trained using 1000 training samples. All results are obtained by averaging over three independent runs. Energy RMSE is given in kcal/mol, while force RMSE is given in kcal/mol/\AA.}
    \begin{center}
    \resizebox{0.6\textwidth}{!}{
    \begin{tabular}{lccrr}
    \toprule
    Model                           & PITC   & PISC   & E                       & F                     \\
    \cmidrule(lr){1-3} \cmidrule(lr){4-5}
    \multirow[c]{4}{*}{PaiNN}       & \xmark & \xmark & 56.62 $\pm$ 2.80       & 12.96 $\pm$ 0.06       \\
                                    & \cmark & \xmark & {\bf 24.60 $\pm$ 0.18} & {\bf 11.51 $\pm$ 0.03} \\
                                    & \xmark & \cmark & 58.30 $\pm$ 2.10       & 13.18 $\pm$ 0.29       \\
                                    & \cmark & \cmark & {\bf 24.53 $\pm$ 0.48} & {\bf 11.43 $\pm$ 0.05} \\            
    \cmidrule(lr){1-3} \cmidrule(lr){4-5}                                   
    \multirow[c]{4}{*}{Equiformer}  & \xmark & \xmark & 54.52 $\pm$ 4.52       & 10.10 $\pm$ 0.00       \\
                                    & \cmark & \xmark & 32.64 $\pm$ 26.48      & {\bf 9.64 $\pm$ 0.03}  \\
                                    & \xmark & \cmark & 48.96 $\pm$ 4.96       & 10.30 $\pm$ 0.06       \\
                                    & \cmark & \cmark & {\bf 20.89 $\pm$ 0.50} & {\bf 9.68 $\pm$ 0.03}  \\            
    \bottomrule
    \end{tabular}}
    \label{tb:Ani1x_ablation_results}
    \end{center}
\end{table}

\subsection{Adversarial Directions for Perturbing Atomic Positions.} The following discusses the dependence of the PIWSL's performance on selecting the vector $\dr$ in \eqref{eq:PILWS-loss} employed to perturb atomic positions. The detailed implementation and setups are provided in \secref{sec:setup}. \Tabref{tb:Ani1x_adv_results} compares the results obtained for a randomly-sampled vector $\dr$ and for the one determined adversarially. The results demonstrate that both approaches improve the performance compared to the baseline without weak supervision, though the results might depend on the employed model.

\begin{table}[h]
    \caption{\textbf{PIWSL's performance dependence on the atomic position perturbation vector.} The numerical values are RMSEs for the energy in kcal/mol and force in kcal/mol/\AA. All results are provided for the ANI-1x data set and models trained using 1000 configurations. \label{tb:Ani1x_adv_results}}
    \begin{center}
    \resizebox{0.7\columnwidth}{!}{
    \begin{tabular}{llccc}
    \toprule
                                    &   & Baseline          & Random (\eqref{eq:adv-random})    & Adversarial (\eqref{eq:adv-direc})    \\
    \cmidrule(lr){1-2} \cmidrule(lr){3-5}
    \multirow[c]{2}{*}{PaiNN}       & E & 56.62 $\pm$ 2.80  & {\bf 24.53 $\pm$ 0.48}            & 33.67 $\pm$ 1.12                      \\
                                    & F & 12.96 $\pm$ 0.18  & {\bf 11.43 $\pm$ 0.05}            & 12.74 $\pm$ 0.14                       \\
    \cmidrule(lr){1-2} \cmidrule(lr){3-5}
    \multirow[c]{2}{*}{Equiformer}  & E & 54.52 $\pm$ 4.52  & 23.16 $\pm$ 0.50                  & {\bf 20.54 $\pm$ 0.21}                 \\
                                    & F & 10.10 $\pm$ 0.00  & 10.03 $\pm$ 0.03                  & {\bf 9.93 $\pm$ 0.04}                  \\
    \bottomrule
    \end{tabular}}
    \end{center}
\end{table}

\subsection{Performance Difference Between First and Second order PIWSL}\label{sec:PIWSL-order-comparison}

In this section, we present a comparison between first- and second-order PIWSL methods. We evaluate a foundation model, MACE-OFF, fine-tuned on the MD17 Aspirin dataset with CCSC(T) labels and the MD22 buckyball catcher data set, as well as PaiNN-GF, a gradient-based force model trained on 1,000 samples from ANI-1x. The results are provided in \Tabref{tb:mace_1st-order_PIWSL}.

Our findings indicate that the second-order term becomes more significant as model performance improves, particularly when force prediction error becomes small. This suggests that the first-order term, which encourages the model to predict the potential energy of neighboring conformations based on a first-order Taylor expansion, plays a crucial role in promoting a smoother potential energy surface. This effect arises because, in MACE and PaiNN-GF models, force predictions are derived as the gradient of the potential energy surface.

\begin{table}[h]
    \caption{\textbf{Performance difference between first and second order PIWSL.} The numerical values are RMSEs (MD17-CC) and MAE (MD22-BB) for the energy in kcal/mol and force in kcal/mol/\AA. MD17-CC denotes MD17 aspirin data with CCSD label, MD22-BB denotes MD22 buckyball catcher data, and ANI-1x (1K) denotes 1000 samples from ANI-1x. \label{tb:mace_1st-order_PIWSL}}
    \begin{center}
    \resizebox{0.8\columnwidth}{!}{
    \begin{tabular}{lllccc}
    \toprule
    Dataset          & Model &  & Baseline & 1st-order PIWSL & 2nd-order PIWSL \\
    \cmidrule(lr){1-3} \cmidrule(lr){4-6}
    \multirow[c]{2}{*}{MD17-CC} & \multirow[c]{2}{*}{MACE-OFF} 
                                & E & 1.21 $\pm$ 0.00 & 0.90 $\pm$ 0.03 & {\bf 0.72 $\pm$ 0.01} \\
                             &  & F & 6.90 $\pm$ 0.01 & 4.59 $\pm$ 0.16 & {\bf 3.77 $\pm$ 0.13} \\
    \cmidrule(lr){1-3} \cmidrule(lr){4-6}
    \multirow[c]{2}{*}{MD22-BB} & \multirow[c]{2}{*}{MACE-OFF} 
                                & E & 1.16 $\pm$ 0.15  & 1.16 $\pm$ 0.28  & {\bf 0.99 $\pm$ 0.05} \\
                             &  & F & 0.35 $\pm$ 0.00  & 0.35 $\pm$ 0.00  & {\bf 0.34 $\pm$ 0.00} \\
    \cmidrule(lr){1-3} \cmidrule(lr){4-6}
    \multirow[c]{2}{*}{ANI-1x (1K)} & \multirow[c]{2}{*}{PaiNN-GF} 
                                & E & 23.57 $\pm$ 0.62 & {\bf 18.62 $\pm$ 0.09} & 18.88 $\pm$ 0.23 \\
                             &  & F & 11.32 $\pm$ 0.08 & {\bf 10.94 $\pm$ 0.01} & 11.01 $\pm$ 0.00 \\
    \bottomrule
    \end{tabular}}
    \end{center}
\end{table}

\end{document}